\documentclass[11pt,english]{article}
\pdfoutput=1
\usepackage{amsfonts, amsmath, amssymb, verbatim, setspace, pdfsync, graphicx,comment}
\usepackage{color}
\usepackage{pdfsync}
\definecolor{darkgreen}{rgb}{0,0.5,0}
\definecolor{darkblue}{rgb}{0,0,0.6}
\definecolor{purple}{rgb}{0.4,0.15,0.21}
\definecolor{black}{rgb}{.2,.2,.2}
\usepackage[colorlinks=true,citecolor=darkgreen,linkcolor=black,urlcolor=darkblue]{hyperref}

\DeclareMathOperator{\p}{\partial}
\DeclareMathOperator{\h}{\theta}

\newcommand{\be}{\begin{equation}}
\newcommand{\ee}{\end{equation}}
\newcommand{\f}{\frac}

\begin{document}
\unitlength = 1mm
\ 
\begin{flushright}
SU-ITP-13/14
\end{flushright}

\begin{center}


{ \LARGE {\textsc{\begin{center}FRW cosmologies and hyperscaling-violating geometries: higher curvature corrections, ultrametricity, Q-space/QFT duality, \\and a little string theory\end{center}}}}

\vspace{0.8cm}

Edgar Shaghoulian

\vspace{.5cm}

{\it Stanford Institute for Theoretical Physics, Stanford University,}\smallskip \\
{\it Stanford, CA 94305-4060, USA}\\

\vspace{1.0cm}

\end{center}

\begin{abstract}

\noindent We analyze flat FRW cosmologies and hyperscaling-violating geometries by emphasizing the analytic continuation between them and their scale covariance. We exhibit two main calculations where this point of view is useful. First, based on the scale covariance, we show that the structure of higher curvature corrections to Einstein's equation is very simple. Second, in the context of accelerated FRW cosmologies, also known as Q-space, we begin by calculating the Bunch-Davies wavefunctional for a massless scalar field and considering its interpretation as a generating functional of correlation functions of a holographic dual. We use this to conjecture a Q-space/QFT duality, a natural extension of dS/CFT, and argue that the Euclidean dual theory violates hyperscaling. This proposal, when extended to decelerated epochs in our own cosmological history like matter or radiation domination, suggests a holographically dual description via RG phases which violate hyperscaling. We further use the wavefunctional to compute Anninos-Denef overlaps and show that the ultrametric structure discovered for de Sitter becomes \emph{sharper} in accelerated FRW cosmologies as the acceleration slows. The substitution $d\rightarrow d_{\textrm{eff}}=d-\theta$ permeates and illuminates the discussion of wavefunctionals and overlaps in FRW cosmologies, allowing one to predict the sharpened structure. We conjecture that the sharpening of ultrametricity is holographically manifested by the growth of the effective dimensionality of the dual theory. We try to find an alternate manifestation of this ultrametric structure by studying the connection of the $\theta\rightarrow -\infty$ background to little string theory.

\end{abstract}

\pagebreak
\setcounter{page}{1}
\pagestyle{plain}

\setcounter{tocdepth}{1}

\tableofcontents

\begin{section}
{Introduction}\label{intro}
\end{section}

Flat FRW cosmologies and hyperscaling-violating geometries have been studied extensively without necessarily utilizing the connection between the two. The relevance of the former is to our cosmological evolution whereas the relevance of the latter is to describing properties of phases of matter via gauge/gravity duality \cite{Maldacena:1997re, Hartnoll:2011fn}. When we refer to hyperscaling-violating geometries, we will always mean ones that are conformal to Lifshitz geometries \cite{Kachru:2008yh}. For dynamical critical exponent $z=1$, an analytic continuation exists that connects hyperscaling-violating geometries to flat, isotropic FRW cosmologies with pure power law scale factor. For general $z$, the continuation connects to an anisotropic Bianchi Type I cosmology; we will only consider isotropic cosmologies but will sometimes consider general $z$ in the context of hyperscaling-violating geometries.

We will begin by studying the effect of higher curvature corrections on both geometries. Our results depend only on the fact that these geometries are scale covariant.\footnote{Unless otherwise stated, we will be talking about a specific form of scale covariance, where the metric can be assigned a scaling weight $\Delta$; see Sections \ref{hyper} and \ref{FRW}.} The fact that they are not scale invariant already suggests that you cannot produce a non-renormalization theorem for them, as exists for AdS, dS, Schr\"{o}dinger \cite{Son:2008ye, Balasubramanian:2008dm}, and Lifshitz spacetimes \cite{Adams:2008zk}. We will see that due to the specific type of scale covariance, the addition of a higher derivative term to a tree-level action that produces an FRW or hyperscaling-violating geometry will lead to equations of motion that cannot be solved with an FRW or hyperscaling-violating metric ansatz, even with renormalized values of parameters. In Lifshitz geometries with matter content respecting the symmetries, for example, one expects only $z$ and $L_{AdS}$ to renormalize, but for the Lifshitz form to be kept. These higher curvature terms can come from both $\alpha '$ (``classical/stringy") and $G_N$ (``quantum") corrections; the source is irrelevant and the conclusion remains the same. We will show that all higher curvature corrections take an exceedingly simple form.\footnote{For pedagogical clarity we emphasize that the results for higher curvature corrections in Section \ref{highercurv} stand completely independently from the rest of the paper and do not inform the discussion in Sections \ref{horizons} - \ref{infinity}. 
}

A distinct set of calculations we will present which emphasizes the analytic continuation between the two geometries is of wavefunctionals and Anninos-Denef field overlaps for a massless scalar on a fixed FRW background \cite{Anninos:2011kh}. We will show that the substitution $d\rightarrow d_{\textrm{eff}}=d-\theta$, which often occurs in the hyperscaling-violating context, is relevant in the case of FRW cosmologies as well. This forms part of the intuition for proposing a Q-space/QFT duality for accelerated FRW cosmologies and arguing that the Euclidean boundary theory violates hyperscaling. The substitution $d\rightarrow d_{\textrm{eff}}$ further lets one immediately predict the extreme structure of correlation functions and Anninos-Denef field overlaps. Specifically, we will see that the two-point function of a massless scalar in the Bunch-Davies vacuum is IR divergent. The power of the divergence depends on the power of the scale factor and can get arbitrarily large; it is this extreme IR structure that leads to ultrametric structure that sharpens as $\theta$ becomes large and negative. Since high-dimensional systems are more often associated with ultrametric structure, we speculate that the growth of the effective dimensionality of the dual theory is the holographic manifestation of the sharpened ultrametricity. Motivated by this sharp structure, we will end by analyzing the $\theta\rightarrow -\infty$ geometry with $z$ finite in the hyperscaling-violating family and comment on its relation to nonlocal or stringy duals.

Section 2 contains the study of higher curvature corrections; it stands alone and is unrelated to the rest of the sections. Section 3 computes perturbative wavefunctionals in an accelerated FRW background in addition to discussing features of these cosmologies which lead to the conjecture of a Q-space/QFT duality. Section 4 utilizes the wavefunctionals of Section 3 to compute cosmological overlaps and illustrates the sharpening of ultrametricity. Section 5 studies the $\theta\rightarrow-\infty$ hyperscaling-violating geometry and can be read alone, though its study is motivated by the results of Section 4. Section 6 contains a discussion of the results and some ideas as to their origin, as well as a few interesting open problems. 
\subsection{Hyperscaling-violating geometries}
In pursuit of exploring phases of matter holographically, geometries have recently been proposed that have both dynamical critical exponent $z$ and hyperscaling violation exponent $\h$ \cite{Huijse:2011ef, Gouteraux:2011ce}. These metrics can be written in the equivalent forms 
\begin{align}
ds_{d+2}^2&=\frac{1}{\tilde{r}^2}\left(-\frac{dt^2}{\tilde{r}^{2d(z-1)/(d-\h)}}+\tilde{r}^{2\h/(d-\h)}d\tilde{r}^2+dx_i^2\right)\label{hyp1}\\
&=r^{2\h/d}\left(-\frac{dt^2}{r^{2z}}+\frac{dr^2+dx_i^2}{r^2}\right),\label{hyp2}
\end{align}
where the latter form makes manifest the fact that these metrics are conformally equivalent to Lifshitz. The parameter $\theta$ breaks scale invariance to scale covariance and maps to the hyperscaling violation exponent in the dual field theory. The Ricci scalar for these geometries is given by $R\sim r^{-2\h/d}$, indicating a curvature singularity at an extreme value of $r$ that depends on the sign of $\h$. Although this and all other curvature invariants are well-behaved at the opposite extreme value of $r$, there exist tidal force singularities there that are precisely analogous to the tidal force singularities of Lifshitz geometries, unless $z=1+\h/d$ with $d/2\leq \theta\leq d$ \cite{Shaghoulian:2011aa,Copsey:2012gw}. Strings are also infinitely excited, just as in Lifshitz \cite{Horowitz:2011gh}, suggesting that the singularity may be upgraded to a full-fledged ``stringularity," although there are a plethora of ways in which this conclusion may be evaded in a real top-down construction.\footnote{See e.g. \cite{Bao:2012yt} for a stressful evasion of a Lifshitz singularity or \cite{Harrison:2012vy, Bhattacharya:2012zu, Knodel:2013fua} for alternative resolutions.}

The resurgence of interest in these geometries is partially due to the fact that, for $\h=d-1$, they exhibit a logarithmic violation of the entanglement entropy \cite{Ogawa:2011bz} when computed holographically \cite{Ryu:2006bv}. This logarithmic violation is prevalent in field theory calculations of fermionic systems \cite{Wolf:2006zzb, Gioev:2006zz}. Due to this and other properties of this metric, such as the specific heat scaling $C\sim T^{1/z}$ with $z\geq 1+\h/d$, it has been proposed that this geometry holographically realizes a non-Fermi liquid (although the low energy spectral density of transverse currents is exponentially suppressed as computed in \cite{Hartnoll:2012wm}). Due to the positive value of the hyperscaling violation exponent, there is a curvature singularity in the UV. This is not problematic because we imagine gluing this metric onto an asymptotically AdS one, which is nonsingular in the UV. To deal with the tidal force singularity in the IR one can either cloak it behind an event horizon, which will allow one to study small but nonzero temperatures, or consider the nonsingular metrics. The latter option is intriguing because $z=3/2$ is the same value of dynamical critical exponent that appears in non-Fermi liquid constructions in $(2+1)$-dimensional field theory \cite{Metlitski:2010pd, Thier:2011gf}.

\subsection{Flat FRW}
FRW geometries are the cornerstone of our description of cosmology. With arbitrary constant curvature spatial slices, we can write the geometries as 
\begin{align}
ds^2_{d+1}=-dt^2+a(t)^2d\Sigma_d^2 \hspace{5mm} \textrm{where } d\Sigma_d^2=\left\{
\begin{array}{c l}      
    dr^2+\sin^2 r\;d\Omega^2_{d-1} & k>0\\
    dr^2+r^2d\Omega^2_{d-1} & k=0\hspace{3mm}.\\
    dr^2+\sinh^2r\;d\Omega^2_{d-1} &k<0\\
\end{array}\right.
\end{align}
$d\Omega^2_{d-1}$ represents the round metric on the unit $(d-1)$-sphere and $k$ measures the curvature of the spatial slices. We shall stick to $k=0$ because only this case has scaling as a conformal isometry. This is made manifest by redefining time and switching to Cartesian coordinates to get 
\be
ds_{d+1}^2=a(\eta)^2\left(\frac{-d\eta^2+dx_i^2}{\eta^2}\right).
\ee
This is different than the usual convention since the $1/\eta^2$ factor is often absorbed into the scale factor, but we write it in this way to mirror the hyperscaling-violating geometry. Notice that for $a(\eta)=1$ we recover de Sitter space and scaling is restored as an isometry. Epochs in our cosmological evolution, like matter domination or radiation domination, are well approximated by various powers of a pure power law evolution for the scale factor.

\begin{section}
{Higher curvature corrections}\label{highercurv}
\end{section}
In the next two subsections we consider the production of hyperscaling-violating and FRW geometries from actions with higher curvature corrections. We will show that these higher curvature terms produce linearly independent tensors at each order in the equations of motion. Our analysis will remain general instead of restricting to a certain theory.

\begin{subsection}
{Hyperscaling-violating geometries}\label{hyper}
\end{subsection}
We consider general theories with both hyperscaling violation ($\theta\neq 0$) and dynamical critical exponent ($z\neq 1$). With the curvature scale set to unity, the metrics are of the form 
\begin{align}
ds_{d+2}^2=r^{2\h/d}\left(\frac{-dt^2}{r^{2z}}+\frac{dr^2}{r^2}+\frac{dx_i^2}{r^2}\right).\label{metric}
\end{align}
This geometry has a symmetry algebra generated by a Hamiltonian $H$, linear momenta $P_i$, and angular momenta $M_{ij}$. These correspond to temporal translation invariance, spatial translation invariance in $x_i$, and rotational invariance between the $x_i$, respectively. In the case where $\h=0$ we reduce to Lifshitz spacetimes, and there exists in addition to these isometries an anisotropic scaling isometry generated by a dilatation operator $D$. For general hyperscaling violation exponent, the dilatation operator given by scaling $t\rightarrow \lambda^z t$, $x_i\rightarrow \lambda x_i$, and $r\rightarrow \lambda r$ (just as in Lifshitz) is a conformal isometry as the metric transforms as $ds^2\rightarrow \lambda^{2\h/d}ds^2$. Even for the nonsingular case $z=1+\h/d$, there exists no isometry corresponding to special conformal transformations (unless $z=1$).

We begin by analyzing the Einstein tensor, which is defined as usual without the cosmological constant: $G_{\mu\nu}\equiv R_{\mu\nu}-\frac{1}{2}Rg_{\mu\nu}$. It appears in Einstein's equation as $G_{\mu\nu}=T_{\mu\nu}$ in our units where $8\pi G_N=1$. We shall soon split the stress-energy tensor to a part which contains the contributions from matter $T_{\mu\nu}^{matter}$ and another which contains the contributions from higher curvature terms $T_{\mu\nu}^{curv}$, with $T_{\mu\nu}=T_{\mu\nu}^{matter}+T_{\mu\nu}^{curv}$. Evaluating the Einstein tensor on the hyperscaling-violating metric ansatz, one gets 
\be
G=G_{\mu\nu}dx^{\mu}dx^{\nu}=\alpha\, \frac{-dt^2}{r^{2z}}+\beta\, \frac{dr^2}{r^2}+\gamma \, \frac{dx_i^2}{r^2}.
\ee
This can be understood by noting that the Einstein tensor is invariant under all the isometries of the metric, in addition to being invariant under Lifshitz scaling. This latter fact follows from the transformation of the Ricci tensor (and of the Ricci scalar multiplied by the metric) under conformal transformations. Indeed, one can show that the most general conserved ($\nabla^\mu G_{\mu\nu}=\nabla^\mu T_{\mu\nu}=0$), symmetric two-tensor invariant under temporal and non-radial spatial translations, non-radial spatial rotations, and Lifshitz scaling must be of this form. Thus, the total stress energy tensor on the right hand side of Einstein's equation must be of this form, even upon including higher derivative terms in the action. To see why this is problematic, imagine that we consider adding curvature squared terms to an action that gives rise to these hyperscaling-violating metrics:
\be
S=\frac{1}{2}\int\sqrt{g}\left(R+L_m(\Phi_i)+c_1 R^2+c_2 R^{\mu\nu}R_{\mu\nu}+c_3 R^{\mu\nu\rho\sigma}R_{\mu\nu\rho\sigma}\right)
\ee
where $c_1, c_2,$ and $c_3$ are field-independent constants. Computing the contribution of these higher curvature terms to the stress energy tensor gives 
\be
T^{R^2}_{\mu\nu}\,dx^\mu dx^\nu= r^{-2\h/d}\left(\alpha_1\frac{-dt^2}{r^{2z}}+\beta_1\frac{dr^2}{r^2}+\gamma_1\frac{dx_i^2}{r^2}\right).
\ee
In fact, adding more higher curvature terms to the action and computing to order $n+1$ in curvature gives a result
\be
T^{R^{n+1}}_{\mu\nu}\, dx^\mu dx^\nu=r^{-2n\h/d}\left(\alpha_n\frac{-dt^2}{r^{2z}}+\beta_n\frac{dr^2}{r^2}+\gamma_n\frac{dx_i^2}{r^2}\right).\label{stressenergy}
\ee
This can be understood in terms of scaling weights; see Appendix \ref{proof2} for a proof. We are considering higher curvature terms that come solely from contractions of the Riemann tensor. In our language, we will refer to a prefactor of $r^{-2n\h/d}$ as having weight $n$. We now write Einstein's equation as 
\begin{align}
\underbrace{\left(R_{\mu\nu}+\frac{1}{2}R g_{\mu\nu}\right)dx^\mu dx^\nu}_{-\frac{\alpha_0 dt^2}{r^{2z}}+\frac{\beta_0 dr^2+\gamma_0 dx_i^2}{r^2}}\hspace{5mm}=\underbrace{T_{\mu\nu}^{tree}\,dx^\mu dx^\nu}_{-\frac{\alpha_m dt^2}{r^{2z}}+\frac{\beta_m dr^2+\gamma_m dx_i^2}{r^2}}+\hspace{3mm}\sum_{n=1}^m\underbrace{T_{\mu\nu}^{R^{n+1}}\, dx^\mu dx^\nu}_{ r^{-2n\h/d}\left(-\frac{\alpha_n dt^2}{r^{2z}}+\frac{\beta_n dr^2+\gamma_n dx_i^2}{r^2}\right)},
\end{align}
where the tree level stress energy tensor (which in this example precisely equals $T_{\mu\nu}^{matter}$) was responsible for producing the necessary metric before adding higher curvature contributions.
Thus, we see that producing a hyperscaling-violating metric in this context requires tuning: since the LHS of Einstein's equation has weight 0, the RHS needs to have weight 0. This can only happen if all the higher weights in $T_{\mu\nu}^{curv}$ cancel order by order (i.e. $\alpha_i=\beta_i=\gamma_i=0$), since they are linearly independent tensors at each order and thus cannot cancel against each other. Of course, if higher curvature corrections are perturbatively small, then these metrics will simply pick up perturbatively small corrections that break the scale covariance. If the hyperscaling-violating metric appears in the IR while the UV remains asymptotically AdS, then the scale covariance will not be exact even without the higher derivative corrections, so these perturbatively small corrections are not problematic.

The situation is worse than it seems in the context of e.g. Einstein-Maxwell-dilaton actions producing electric solutions. The gauge-kinetic coupling is vanishing in the deep IR, which means that $\alpha'$ corrections are becoming important. But these corrections include precisely the higher curvature terms which increase in importance and preclude the existence of a hyperscaling-violating metric as a solution to the metric equation of motion! Thus the geometry can be deformed much more than perturbatively (but see Section \ref{loop} for a possible way out).

The reader may at this point challenge the form of these higher curvature terms. Namely, if they had field-dependent prefactors, it is possible to change the weights on the RHS of Einstein's equation since the fields can carry weights of their own. In the simplest case this would require at each order $n+1$ (in the action) in curvature to have a field-dependent prefactor $F_i[\Phi_j]$ that has weight $-n$ to cancel the curvature weight of $n$ (in the stress energy tensor). This is a logical possibility and represents the simplest way of evading our conclusions (it is relatively straightforward to use this approach to construct various higher curvature actions which have hyperscaling-violating geometries as a solution). One can also imagine other suppressed operators entering the picture and providing the requisite cancellation between $T_{\mu\nu}^{curv}$ and $T_{\mu\nu}^{matter}$, but again this cancellation would have to work exactly at every order to maintain the scale covariance.

Another way to evade these conclusions is to include sufficiently complicated matter or interactions, which may come naturally from other suppressed operators, to provide the necessary terms in $T_{\mu\nu}^{matter}$ to offset the contribution of $T_{\mu\nu}^{curv}$. Of course, the higher one goes in derivatives, the more matter or interactions one needs to put into the action and tune accordingly to get the required cancellations. This is what is sometimes done when producing FRW metrics in higher derivative gravity, as discussed in Section \ref{FRW}.

\subsubsection{A potential exception}\label{loop}

We have seen that, unsurprisingly, no non-renormalization theorem holds for the hyperscaling-violating metrics as does for Lifshitz and Schr\"{o}dinger metrics. In the case of asymptotically AdS space, we can ask about the existence of the hyperscaling-violating metrics deep in the bulk, i.e. as the IR theory of our RG flow. As we showed earlier, all curvature invariants behiave as $r^{-2n\h/d}$ for some $n\geq 0$. Notice also that the IR is always at large $r$ in these coordinates for the NEC-satisfying range we care about, $z\geq 1+\theta/d$. Thus, as long as $\h>0$, all higher derivative corrections will vanish in the IR. The argument of the previous subsection therefore would not apply.\footnote{Of course one can also say it breaks down as $r\rightarrow 0$ for $\h<0$; this may be interesting as in this case you may have a stable UV boundary with no curvature singularity (as opposed to when $\h>0$), robust against higher derivative corrections.} 

In a top-down construction, however, the situation is a bit more complicated. The dilaton of the effective theory (which we here take to be the Einstein-Maxwell-dilaton theories originally used to produce these geometries) needs to be connected to the string theory dilaton. If, for example, the string theory dilaton is precisely related to the effective field theory dilaton as below, then the higher curvature corrections carry $\alpha'$ prefactors which are related to the dilaton as follows:

\begin{align}
M_s\sim g M_P & \sim 1/\sqrt{\alpha'}\,,\hspace{5mm} \textrm{where } g\sim e^{-a \phi/2}\\
&\implies \alpha'\sim e^{a\phi},
\end{align}
where we imagine $M_P$ is fixed and $1/g^2$ is given by the coefficient of the gauge-kinetic term in the action. This is why higher derivative terms become important at small $g$, since the corrections are accompanied by $\alpha'$ factors which are getting large. In this setup (or any one where one knows the precise relationship between the effective theory dilaton and the string theory dilaton), one would have to compare the growth of $\alpha'$ with the vanishing of the curvature invariants. However, it is possible that the relation between string theory and effective theory dilatons allows for the higher curvature terms to vanish in the IR.\footnote{While assuming this we can also assume that the higher derivative terms, e.g. $\alpha' F^4$, also vanish, since in these constructions $F^2\rightarrow 0$ in the IR.} As stressed by \cite{kachru}, given that the string mass is getting light, one should expect on general grounds that the description in the IR is still breaking down in one way or another. If  the exception stated above is realized, though, then an explicit example of what sorts of corrections change the IR geometry in the nonsingular case is lacking, since higher curvaure terms remain small, the metric is regular, and probe string excitations do not diverge.

\subsection{FRW geometries}\label{FRW}
Let us now turn back to the flat FRW metric
\be
ds_{d+1}^2=-dt^2+t^{2q}dx_i^2
\ee
with $q>0$ so that it is an expanding spacetime. The null energy condition is trivially satisfied as long as $q\geq 0$. The coordinate ranges are $x_i\in \mathbb{R}$ and $t\in \mathbb{R}^+$. The ``Big Bang" is at $t=0$, where the metric has a curvature singularity. To make manifest the scale covariance, define
\be
\f{dt}{t^q}=d\eta\implies \eta=\f{t^{-q+1}}{1-q}.
\ee
Then we find
\begin{align}
ds_{d+1}^2=N\;\eta^{\f{2q}{1-q}}(-d\eta^2+dx_i^2)\longrightarrow ds_{d+1}^2=|\eta|^{\f{2q}{1-q}}\;(-d\eta^2+dx_i^2)
\end{align}
where we have rescaled a $q$-dependent constant $N$ into the coordinates. This rescaling obscures the Lorentzian signature of the metric, so we have taken the absolute value of $\eta$ to make this fact manifest. The point $q=1$ is special since our transformation breaks down.\footnote{In the $q=1$ case the scale-covariant metric becomes $e^{2\eta}(-d\eta^2+dx_i^2)$. The hyperscaling-violating cousin of this metric corresponds to $\theta\rightarrow -\infty$ with $z$ fixed and finite. This geometry will be studied in Section \ref{infinity}.} Defining $q=\f{1-d+\theta}{\theta}$, we get
\be
ds_{d+1}^2=\eta^{\f{2\theta}{d-1}}\left(\f{-d\eta^2+dx_i^2}{\eta^2}\right),\label{frwanal}
\ee
which for $q>0$ and $d>1$ means $\theta \in (-\infty,0)\cup  (d-1,\infty) $. We have dropped the absolute value on $\eta$, but it is implied. 

Notice that from here it is manifest that flat, isotropic FRW geometries are analytic continuations of hyperscaling-violating geometries with $z=1$. Up to the shift in the definition of $d$ and an overall sign, the continuation from (\ref{metric}) to (\ref{frwanal}) is given by $x_i\rightarrow ix_i$, $t\rightarrow x_{i+1}$, $r\rightarrow \eta$. To get the mostly plus convention of (\ref{frwanal}) one simply needs to appropriately reintroduce a curvature scale and analytically continue it, as is done in going from AdS in Poincar\'e coordinates to dS in flat slicing. Notice that in the case of $\h=0$ one often simply continues the radial and time coordinates of the AdS metric, but in this case a different continuation is necessary due to the arbitrary power on the radial coordinate disallowing the continuation $r\rightarrow i\eta$. Analytic continuations between domain wall backgrounds and cosmologies have also been used for different purposes in \cite{Skenderis:2006jq, Skenderis:2006fb, Skenderis:2007sm, Kiritsis:2013gia}.

In the context of FRW cosmology, the parameter $\h$ does not yet carry the interpretation as the hyperscaling violation exponent in a purported dual theory (more on this in Section \ref{holo}); it is simply chosen here to illustrate the mathematical analogy with the previous geometries. Namely, this geometry has spatial translations and rotations as isometries and isotropic scaling as a conformal isometry (the isometry is broken by $\h$). More importantly, the same statement about curvature invariants that was true for the hyperscaling-violating geometries remains true for these geometries: the Ricci scalar scales as $R\sim \eta^{-\frac{2\h}{d-1}}$ and contributions to Einstein's equation at order $n+1$ are given by 

\be
T_{\mu\nu}^{R^{n+1}}\, dx^\mu dx^\nu= \eta^{-\frac{2n\h}{d-1}}\left(\alpha_n\frac{-d\eta^2}{\eta^2}+\beta_n\frac{dx_i^2}{\eta^2}\right).
\ee
So we seem to have the same argument as before for the non-genericity of such solutions in higher curvature gravity. Sound the trumpets \cite{Adams:2008zk}.

Actually, not just yet on the trumpets \cite{Adams:2008zk}. There is a large literature regarding the production of spatially flat FRW universes with power law scale factor \cite{Sotiriou:2008rp, DeFelice:2010aj, Nojiri:2006ri, Fay:2007uy, Capozziello:2006dj, Jaime:2012gc, Faraoni:2008mf, GilMarin:2011xq,  Nojiri:2003ft, Ohta:2004wk, Maeda:2004hu} in higher curvature gravity. Specifically, most of these papers discuss the phenomenological viability of $f(R)$ gravity, where the higher curvature terms come solely from powers of the Ricci scalar. This phenomenological viability requires reproducing epochs in our universe like matter and radiation domination, which correspond to specific nonzero values of $\theta$. Are these models simply appropriately tuned, as discussed in the previous subsection?

Yes and no. These models are not considering arbitrarily many higher curvature terms. In fact, many truncate to just two or three powers of the Ricci scalar in $f(R)$. For example, \cite{Brevik:2006wa} finds it necessary to use two fluid sources to produce two powers of the Ricci scalar. Furthermore, many of these papers reproduce matter and radiation domination in regimes where a single power of the Ricci scalar dominates the other powers in the action (see \cite{Nojiri:2008nk} as an example); the other powers give small corrections as discussed before. These seem to be the simplest ways to `evade' our argument.

It is also interesting to ask what happens to the argument when one uses the fact that $f(R)$ gravity can be rewritten as a self-interacting scalar minimally coupled to Einstein gravity (see e.g. \cite{Nojiri:2012zu}). The proof obtains by introducing an auxiliary scalar field $\phi$ and writing the action of $f(R)$ gravity as 
\be
S=\int d^4x\sqrt{-g}\;[f'(\phi)(R-\phi)+f(\phi)].
\ee
Varying the action with respect to the auxiliary field $\phi$ gives an equation solved by $\phi=R$, which reproduces the action of $f(R)$ gravity. Performing a conformal rescaling $g_{\mu\nu}\rightarrow e^{\phi}g_{\mu\nu}$, where $\tilde{\phi}=-\textrm{ln }f'(\phi)$, gives 
\be
S=\int d^4x\sqrt{-g}\left(R-\frac{3}{2}\p_{\mu}\tilde{\phi}\p^{\mu}\tilde{\phi}-V(\tilde{\phi})\right)
\ee
for $V(\tilde{\phi})=\phi/f'(\phi)-f(\phi)/f'(\phi)^2$. Arbitrarily many higher curvature terms get repackaged into arbitrarily many terms in the potential of different weights. So, as one should expect, the same problem exists here.

\begin{section}
{FRW horizons, holography, and wavefunctionals}\label{horizons}
\end{section}
\noindent Having discussed the usefulness of the specific form of scale covariance of FRW cosmologies and hyperscaling-violating geometries, we now move on to a set of conceptually distinct calculations which utilize the analytic continuation between the two geometries as a guiding principle. Namely, we will see that the substitution $d\rightarrow d_{\textrm{eff}}=d-\theta$ is a useful heuristic even in the case of FRW cosmologies.\\
\subsection{Horizons}
\begin{figure}[tb]
\centering
\includegraphics[scale=0.65]{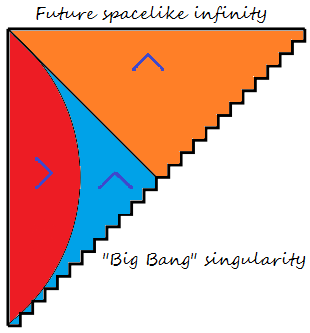}
\caption{The Penrose diagram for accelerated FRW cosmologies. For an observer at the origin, the red region represents the interior of the apparent horizon, the blue region represents the region between the apparent horizon and the cosmological horizon, and the orange region represents the region outside of the cosmological horizon. The Bousso wedges indicate that one can project all the bulk data onto an optimal screen at future spacelike infinity.}
\label{penrose}
\end{figure}

Consider again the FRW metric with flat slicing: 
\be
ds^2=-dt^2+t^{2q}dx_i^2\;,
\ee
with $q>0$ so that it is an expanding spacetime. Notice that for $q>1$ ($-1<w<-1/3$ for a perfect fluid equation of state $p=w\rho$) we have $a''/a>0$, so the expansion is accelerating. These spaces are known as Q-space and have inspired phenomenological interest in the recent past, before our late-time accelerated expansion was shown to be so closely approximated by a cosmological constant. In this case since $\eta=-t^{-c}$ with $c>0$, we see that $\eta$ will be ranging from $-\infty$ to $0$ while $t$ ranges from $0$ to $\infty$. This coordinate range mirrors that of de Sitter and corresponds to $\theta<0$ with $d>1$. For $q=1$ the expansion is inertial ($a''/a=0$) and for $q<1$ it is decelerating. Thus, the $q>1$ spacetimes must have cosmological event horizons. We can compute the proper radius of the event horizon as usual:
\be
\ell = t^q \int_t^\infty \frac{dt'}{(t')^q}=\f{t}{q-1}=\f{t\,\theta}{1-d}\;,
\ee
which is manifestly positive in the regime we are interested in. Notice that unlike the de Sitter case this horizon \emph{increases} in time and diverges at the future spacelike boundary. However, the expansion of space is faster than the growth of the horizon as long as $q>1$, so the full Penrose diagram has cosmological horizons at future spacelike infinity. The apparent horizon is different and found to be
\be
\ell_{AH}=1/H=t/q,\qquad H=a'/a.
\ee
Thus, the apparent horizon is always inside the event horizon. A Penrose diagram representing this state of affairs, with associated Bousso wedges, was presented in \cite{Hellerman:2001yi} and is reproduced and embellished in Figure \ref{penrose}. Unfortunately, no relation along the lines of $s\sim T^{d-\h}$ seems to apply to these horizons, although for small $\h$ they are thermodynamical in the usual sense \cite{Bousso:2004tv}. The marginal case $w=-1/3$ has an apparent horizon but no event horizon.

\subsection{Wavefunctionals}
We would now like to compute a wavefunctional for a scalar field in a fixed FRW background. The vacuum state will be defined adiabatically; namely, we will demand that the UV Fourier modes are purely positive frequency \cite{Hartle:1983ai, Bunch:1978yq}. We will see what this means in a moment. Consider a massive scalar field minimally coupled to the $(d+1)$-dimensional FRW background $a(\eta)^2(-d\eta^2+dx_i^2)$:
\be
S=-\f{1}{2}\int d^{d+1}x\sqrt{g}(\p_\mu\varphi\p^\mu\varphi+m^2\varphi^2).
\ee
Fourier decomposing our field $\varphi(x,\eta)=\int d^3 k \;\varphi_k(\eta)\; e^{ikx}$ we get 
\be
S=\f{1}{2}\int d\eta\int d^d k\; a(\eta)^{d-1}\, \left(\dot{\varphi}_k\dot{\varphi}_{-k}-(k^2+m^2)\varphi_k\varphi_{-k}\right).
\ee
The equation of motion is given by 
\be
\f{1}{\sqrt{g}}\p_\mu(\sqrt{g}g^{\mu\nu}\p_\nu\varphi)=m^2\varphi\;.
\ee
We can Fourier expand and solve this equation for $m=0$ (general $m$ only admits an analytic solution for special cases like $\theta=0$ (dS), $\theta=d-1$ (flat space), and $\theta=(d-1)/2$), obtaining
\be
\varphi_k(\eta)=A_{\vec{k}}^+ \, v_k(\eta)+A_{\vec{k}}^- \,v_k^*(\eta),\qquad v_k(\eta)=(-\eta)^{\f{d-\theta}{2}}H_{\f{d-\theta}{2}}^{(1)}(k\eta),\label{saddle}
\ee
where some choice of $k$-dependent overall normalization needs to be made. Notice that $\theta=0$ reproduces the usual de Sitter answer, while in general $\theta$ acts as an effective shift in the dimensionality. While the Hankel part of the answer looks like the case for a massive scalar in de Sitter, with $\h\sim 2\Delta_-$, notice that the time dependence sitting outside is modified. The Fourier modes of the scalar are instead identical to those of a higher-dimensional massless scalar in de Sitter. This leads to a solution at late times which behaves as
\be
\varphi_k(\eta)=\eta^{\tilde{\Delta}^+}(f(\vec{k})+\mathcal{O}(\eta^2))+\eta^{\tilde{\Delta}^-}(g(\vec{k})+\mathcal{O}(\eta^2))\label{latetime}
\ee
with
\be
\tilde{\Delta}^+=0,\qquad \tilde{\Delta}^-=d-\theta.
\ee
The tilded variables represent mustachioed impostors for conformal weights; more on this shortly. In the case of FRW, it is not appropriate to read off the scaling of the correlators from the field falloffs (i.e. the correlator $\langle \phi(x)\phi(y)\rangle$ does not scale as $|x-y|^{-\tilde{\Delta}^+}$ or $|x-y|^{-\tilde{\Delta}^-}$), whereas in the case of de Sitter this is simply a shortcut which works due to the scale invariance. Although we can similarly apply a scaling argument to get the answer, let us proceed from the wavefunctional point of view. 

Moving forward, we will assume $\theta<0$ since it is the accelerated FRW cosmologies we care about. Implementing the Bunch-Davies vacuum, we require that $\varphi_k(\eta)\sim e^{ik\eta}/\sqrt{k}$ as $k/a(|\eta|)=k\,|\eta|^{\f{d-1-\theta}{d-1}}\rightarrow \infty$. In words, this states that in the limit that the physical wavelength is small in units of the curvature scale (here set to $1$), the field mode should behave as if it were in the Minkowski vacuum. Similar to the usual Bunch-Davies condition in de Sitter, it can be understood by analytic continuation from a formal regularity condition in the Euclidean hyperscaling-violating geometry. However one defines it, this fixes $A_k^-=0$. Writing the saddles as 
\be
\varphi_{\vec{k}}(\eta)=\phi_{\vec{k}}\;\f{v_k(\eta)}{v_k(\eta_c)}
\ee
and evaluating the wavefunction $\Psi\sim e^{iS}$ as a function of the boundary condition at $\eta_c$ we find
\be
\Psi\sim \exp\left[i\int d^d k \, \eta_c^{1-d+\theta}\,\f{v_k'(\eta_c)}{v_k(\eta_c)}\phi_{\vec{k}}\phi_{-\vec{k}}\right].\label{wavefunc}
\ee
\subsection{Correlation functions and holography}\label{corr}
Treating the wavefunctional (\ref{wavefunc}) as a potential generating functional of correlation functions, $\Psi_{HH}=Z_{QFT}$, one obtains
\be
\langle \mathcal{O}(\vec{x})\mathcal{O}(\vec{y}) \rangle \sim \int d^d p_1 d^d p_2\;e^{i\vec{p}_1\cdot \vec{x}}e^{i\vec{p}_2\cdot \vec{y}}\;\f{\delta^2 \Psi}{\delta \phi_{\vec{p}_1} \delta \phi_{\vec{p}_2}}\Bigg| _{\phi=0}\sim \;\f{1}{|\vec{x}-\vec{y}|^{2d-\theta}}\; ,\label{differentiate}
\ee
where we have simply analytically continued the two-point function from the hyperscaling-violating geometry, computed in \cite{Dong:2012se}. We have only kept the spatial dependence, not the various factors of $i$ and $-1$ which can enter and indicate a violation of reflection positivity. Notice that from here one would usually identify the ``conformal" weight as $\Delta^+=d-\theta/2$. The scale covariance of the two-point function (\ref{differentiate}), usually indicative of a scale-invariant field theory description, can be explained by a ``generalized" scale invariance, which will be discussed shortly.

The cosmological correlator can also be computed with the wavefunctional. Focusing just on the momentum dependence and thus ignoring geometric factors, we have
\begin{align}
\langle \phi_k\phi_{k'}\rangle &\sim \frac{\int \mathcal{D}\phi \;|\Psi[\phi]|^2\;\phi_k\phi_{k'}}{\int \mathcal{D}\phi \;|\Psi[\phi]|^2} \sim  \frac{\int \mathcal{D}\phi \;e^{\int d^d \tilde{k}\, \tilde{k}^{d-\theta}\phi_{\tilde{k}}\phi_{-\tilde{k}}}\;\phi_k\phi_{k'}}{\int \mathcal{D}\phi \; e^{\int d^d \tilde{k}\, \tilde{k}^{d-\theta}\phi_{\tilde{k}}\phi_{-\tilde{k}}}}\\
&\sim k^{-d+\theta}\, \delta(k-k')
\end{align}
where the momentum dependence is extracted by rescaling $\phi_k\rightarrow \phi_k/k^{(d-\theta)/2}$. This is IR-divergent \cite{Ford:1977in}  and does not have a well-defined Fourier transform without regularization. One possible scheme is to formally integrate by parts to define the transform, as is often done in differential regularization schemes \cite{Freedman:1991tk, Freedman:1992gr}  used to handle ultraviolet singularities:
\begin{align}
\langle \phi(x)\phi(y)\rangle &\sim \int d^d k\, d^dk' \;k^{-d+\theta}\,\delta(k-k')\,e^{ikx}e^{ik'y}\sim \int dk \;k^{-1+\theta}\,e^{ik(x-y)}\label{lineone}\\
&\sim \int dk \,e^{ik(x-y)} \,\f{d^{-\h}}{dk^{-\h}}\,\frac{1}{k}\sim \int dk \,\f{d^{-\h}}{dk^{-\h}}\,e^{ik(x-y)} \,\frac{1}{k}\label{linetwo}\\
&\sim |x-y|^{-\h}\log|x-y|.\label{linethree}
\end{align}
The derivative relationship for $k^{-1+\theta}$ used in going from (\ref{lineone}) to (\ref{linetwo}) was true for negative integer $\theta$, but in (\ref{linethree}) we can analytically continue $\theta$ to any negative real number. This regularization is only meant as a cartoon. However one wishes to discuss the two-point function, it is clear that it has a strong spatial IR divergence, which we imagine regulating with a spatial IR cutoff. A late-time cutoff is not necessary since the field does not grow with time. The UV properties should be well-behaved due to the Bunch-Davies condition, and the perturbation theory should be in control given the field falloffs (\ref{latetime}). In other words, the stress-energy of the scalar can be kept small relative to the background stress-energy. In the language of \cite{Harlow:2011ke}, this would be the relevant correlation function for the ``extrapolate" dictionary, whereas (\ref{differentiate}) would be relevant for the ``differentiate" dictionary. Notice that, since $\theta<0$, the cosmological correlator grows with distance, violating the cluster decomposition theorem. The violation can be made arbitrarily severe as $\theta\rightarrow -\infty$. For $\theta=0$ the expression is understood as the usual logarithmic growth in de Sitter space. One possible interpretation of this result is that the Bunch-Davies state does not define a suitable vacuum state.\footnote{Massless higher spin fields in de Sitter space have similar strong divergences, with the strength of the divergence increasing with the spin. This needs to be properly understood in the context of the Vasiliev/$Sp(N)$ duality \cite{Anninos:2011ui, Anninos:2012ft,Anninos:2013rza,Ng:2012xp,Das:2012dt,Karch:2013oqa,Banerjee:2013mca}, although there gauge-invariant operators may require a sufficient number of derivatives to tame the strong growth. I am indebted to Daniel Roberts and Douglas Stanford for discussions about the growth of cosmological correlators.}  However, as is the case for disordered systems like the Sherrington-Kirkpatrick model, or even more simply the Ising model below $T_c$, we assume that our vacuum state can be decomposed into pure states which \emph{do} satisfy cluster decomposition. We can then study the state space in detail by computing overlaps between such pure states without ever explicitly performing the decomposition. We will perform these calculations in Section \ref{cosmiccluster} and see that the vacuum state defined above leads to sensible overlaps. Beyond linearized order, adding self-interaction and computing loops will help to understand the infrared effects of this spacetime, and it would diagnose whether the perturbative expansion about the free-field fixed point is breaking down. Since the intuitive picture of modes freezing and classicalizing outside the horizon should carry through from de Sitter, one can also see if Starobinsky's model of stochastic inflation \cite{Starobinsky:1982ee} can be quantitatively generalized to Q-space.

\subsection{The Q-space/QFT correspondence}\label{holo}
Although we most care about interpreting FRW cosmologies as part of an RG flow that represents a cosmology in dS/CFT \cite{Strominger:2001pn, Strominger:2001gp,Witten:2001kn,Maldacena:2002vr}, in which case we would state that the operator $\mathcal{O}$ dual to the massless bulk scalar has obtained an anomalous dimension of $\theta/2$, we can also consider the possibility of holographically describing one of these accelerated FRW spacetimes on its own. Referencing the Penrose diagram of Figure \ref{penrose}, we see that a holographic screen is obtained by projecting along the Bousso wedges to the future spacelike infinity. This represents an optimal screen and encodes the entire bulk spacetime \cite{Bousso:2002ju, Bousso:1999dw, Bousso:1999xy}. Thus, entertaining the possibility of holographically describing accelerated FRW by a field theory living at future spacelike infinity seems well-grounded. We conjecture a natural extension of the dS/CFT correspondence to this background and propose $\Psi_{HH}=Z_{QFT}$, which was used to compute (\ref{differentiate}). Whether this can extend to a non-perturbative definition of the bulk theory is completely unclear at this point. The dual theory would be Euclidean, defined on $\mathbb{R}^{d}$, and in general not reflection positive, although the range of masses for a bulk scalar that keep the boundary conformal dimension real probably increases as $\theta$ gets more negative. Writing the bulk metric as
\be
ds^2=e^{-2t\h/(d-1)}\left(-dt^2+e^{2t}dx_i^2\right),
\ee
we see that a bulk scale transformation $t\rightarrow t+\lambda$, $x_i\rightarrow e^{-\lambda}x_i\hspace{1mm}$  generates a time translation in the bulk and a scale transformation in the boundary. Just as in dS/CFT we therefore identify the holographic time coordinate as corresponding to the energy scale in the boundary theory. The RG flow goes from the IR at $t=-\infty$ to the UV at $t=\infty$. The Big Bang singularity that exists in these spacetimes would naively correspond to a mass gap in the IR, but here we keep unspecified what relevant operators are switched on in the UV, i.e. we leave the IR asymptotics general and consider a holographic theory of asymptotically Q-space.

We further conjecture that the dual theory violates hyperscaling due to the modified scaling dimensions in the boundary partition function, which come from the scaling weight of the metric. The scaling of the metric maps to the anomalous scaling of the stress-energy tensor, which feeds into the other anomalous scalings characteristic of hyperscaling violation. This is allowed dimensionally since there is an additional scale $\ell_Q$ in the metric that we have suppressed. The stress-energy tensor we are referring to is the one given by the Brown-York construction \cite{Brown:1992br}, although other potentially sensible proposals which connect the boundary stress-energy tensor to the bulk graviton will likely give similar results. To proceed with the Brown-York construction and avoid the difficult task of holographic renormalization, we will work at a finite $\eta_c$ slice and not yet concern ourselves with taking $\eta_c\rightarrow 0$. If we imagine that the construction applies at such a slice, then one obtains
\be
\langle T_{ij}\rangle \sim \f{h_{ij}}{\eta_c^{d-\theta}}\;,\label{brownyork}
\ee
with $h_{ij}$ proportional to the induced metric at $\eta_c$. This is just an analytic continuation of the asymptotically AdS case \cite{Dong:2012se}. This anomalous scaling of the momentum density $ [T_{ii}]= d-\h$ is the essence of hyperscaling violation. It suggests that in the cosmological scenario we should take seriously the idea that the effective dimensionality of the dual is given by $d_{\textrm{eff}}=d-\theta$. In a microscopic realization of the Q-space/QFT duality, we can imagine putting the dual theory at finite density by considering a bulk $U(1)$ gauge field. In other words, we source a global $U(1)$ current in the boundary theory with a chemical potential $\mu$. There should then exist a hyperscaling violating regime where the free energy density $f\sim \tau^{d-\h}$, with $\tau\sim \mu^{-1}$ the only length scale in the problem besides $\ell_Q$. This is maybe a more standard way of observing hyperscaling violation, but such an effect can ultimately be traced back to an anomalous scaling of the stress-energy tensor. A pedagogical example of such a zero temperature quantum phase transition was constructed in the context of AdS/CFT in \cite{Ammon:2012je}.

Like incarnations of dS/CFT that do not restrict to the static patch, Q-space/QFT is at heart a theory of metaobservables inaccessible to a local observer. As has been stressed before in the literature, we can be considered to be metaobservers of our previous inflationary epoch \cite{Maldacena:2002vr, Anninos:2012qw}. If that era is correctly described by the slow-roll inflationary paradigm, then Q-space with $\mathcal{O}(\varepsilon)$ violation to hyperscaling is a more accurate model than de Sitter. The small parameter $\varepsilon$ is just the slow-roll parameter. Similarly, our late-time cosmological expansion may be modeled with an $\mathcal{O}(.1)$ violation of hyperscaling and fit the current data just as well as a de Sitter phase \cite{Ade:2013zuv}, although such a scenario may be considered theoretically less motivated than a genuine cosmological constant. 

Though we expect holography for an isolated, decelerated FRW cosmology to behave differently due to the distinct conformal structure, decelerated phases like matter or radiation domination should fit nicely into the RG flow of an asymptotically dS/CFT or Q-space/QFT. They would describe phases of a holographic dual with $\mathcal{O}(1)$ violations of the hyperscaling hypothesis. In our universe, according to our parametrization of $\theta$, radiation domination corresponds to $\theta=4$ and matter domination corresponds to $\theta=6$. Curiously, curvature domination corresponds to infinitely large hyperscaling violation. In general, accelerated expansion leads to $\theta\leq 0$ and decelerated expansion gives $\theta\geq d-1$. The range $0<\theta<d-1$ corresponds to crunching cosmologies which violate the null energy condition. Hyperscaling violation is only a precise notion if we assume a single source of energy density dominates; if the scale factor is not a pure power law, then the metric loses its distinguished scaling weight. In our universe, the dominance of a single source of energy density is approximately true for various epochs and violated in transition regions. The RG flow of a holographic dual to our universe therefore has regimes of approximate hyperscaling violation with transition regions without any well-defined scaling. The monotonicity of the RG flow, if true for these potentially non-unitary theories, likely corresponds to the monotonic growth of the horizon.

As mentioned previously, the scale-covariant form of the two-point function for the operator dual to a massless scalar, which usually indicates an underlying CFT description, can be understood as following from a ``generalized" scale invariance, under which $\eta\rightarrow \lambda \eta$, $x_i\rightarrow\lambda x_i$ and $\ell\rightarrow \lambda \ell$. This leaves the bulk metric invariant but does not preserve the Hilbert space since $\ell$ gets rescaled. Nevertheless, it can be used to constrain the form of correlation functions. This type of generalized scale invariance was first studied in the case of matrix models for $D0$-branes in \cite{Jevicki:1998yr} and generalized to non-conformal $Dp$-branes in \cite{Jevicki:1998ub}. It was further discussed in \cite{Boonstra:1998mp, Kanitscheider:2008kd}.

Although these spacetimes have fewer symmetries than de Sitter, there is a sense in which quantum gravity in such a spacetime may be simpler to analyze: as computed in Section \ref{horizons}, the horizon size diverges near the future boundary. Similarly, the entropy passing through an observer's past light sheet, as computed by the Bousso covariant entropy bound, diverges at late times. This admits the possibility of a precise holographic dual, as discussed in \cite{Witten:2001kn, Bousso:2004tv, Harlow:2010my}, although the issue of imprecision may only be relevant for holographic descriptions of local observers, which is not what we are performing here. Of course, one can write down and analyze a coordinate patch relevant for a local observer, but such a patch would be time-dependent \cite{Hellerman:2001yi}.

Concrete calculations that would help get a handle on this Q-space/QFT correspondence include analyzing higher spin fields in the bulk or calculating an asymptotic symmetry group. A bulk gauge field supplied by the scale-invariant Maxwell action, for example, will lead to a Bunch-Davies wavefunctional identical to that of de Sitter, as calculated in Appendix D of \cite{Anninos:2013rza}. In the calculation of an asymptotic symmetry group one would have to decide how to treat the radiation flux, which can in principle reach $\mathcal{I}^+$ \cite{Anninos:2011jp}. In addition, one needs a sensible prescription for the asymptotics of any non-gravitational sector necessary to source the FRW spacetime. Finally, another calculation which will shed light on Q-space directly and hopefully the Q-space/QFT duality is that of field overlaps in cosmology, which we now turn to.

\begin{section}
{Cosmic clustering of accelerated FRW}\label{cosmiccluster}
\end{section}
\noindent In a remarkable recent paper \cite{Anninos:2011kh}, Anninos and Denef discovered an encoding of the hierarchical tree-like structure of de Sitter space in the Bunch-Davies vacuum. The authors imported the analysis of the state space of spin glasses \cite{spinglassbook, LNF-79-31-P, parisi2, 198330, 107911, 244332} and tailored it to the study of quantum fields on a fixed de Sitter background. The philosophy has been laid out most clearly in \cite{Anninos:2011kh, Denef:2011ee, diothesis}, and we refer the reader to these works for more details, though we will summarize the key points as we go along. The idea is to split up the Bunch-Davies state for the massless scalar, which does not satisfy cluster decomposition, into a sum of ``pure" states which do. Overlaps between these states, which we will define below, then reveal the extreme state structure: the one-point distance distribution is a Gumbel distribution and the three-point distance distribution reveals ultrametric structure. Related distance distributions were considered in \cite{Benna:2011as} and the extension to massive scalars in de Sitter was considered in \cite{Roberts:2012jw}. The latter paper found that the extreme state structure weakens but persists until $\Delta_- = d/4$. 

In what follows, we will compute such overlaps in the case of a massless scalar field in the Bunch-Davies vacuum in an accelerated FRW cosmology. We will find that the structure uncovered in \cite{Anninos:2011kh} becomes \emph{sharper} in the context of accelerated FRW cosmologies. All statements we make about relative sharpness will be referencing the distance function as defined in \cite{Anninos:2011kh} and used in (\ref{dist1}), although given the more extreme IR behavior of accelerated FRW cosmologies relative to de Sitter, we expect that other ``natural" distance functions will give the same results.
\\
\subsection{Distance overlaps}
To maximally overlap with previous literature, we will place an IR cutoff of comoving size $L$ such that $x_i \sim x_i+L$ and write our results from the previous sections as follows:
\be
\mathcal{P}[\phi]=|\Psi[\phi]|^2\propto e^{-2\sum_{\vec{k}}\;\beta_k|\phi_{\vec{k}}|^2},\qquad \beta_k\equiv \textrm{Re}\left[-\frac{i}{2}\;a^{d-1}L^d(\log v_k)'\right],\label{wavefunctional}
\ee
where the IR cutoff has changed what was originally an integral into a sum. This is the wavefunctional as a functional of field distributions $\phi_{\vec{k}}$ at $\eta_c$. It is evaluated by inputting the complex saddle found in (\ref{saddle}) to evaluate the path-integral definition of the ground state wavefunctional. Notice that when written in terms of $\eta$, most of the change in (\ref{wavefunctional}) from the de Sitter expressions can be accounted for by $d\rightarrow d-\theta$. The sum over momenta, however, goes over $d$ and not $d-\theta$ dimensions, and this is the crucial difference (otherwise the expressions would be identical upon identifying $d-\theta\rightarrow \tilde{d}$). 

Given two field configurations $\phi_1(x_i)$ and $\phi_2(x_i)$ on a given time slice, the distance between them is defined as 
\be
d_{12}=\frac{1}{L^d}\int d^d x (\hat{\phi}_1(x_i)-\hat{\phi}_2(x_i))^2.\label{dist1}
\ee
The hatted variables correspond to taking the original field distribution, convolving it with a window function over a size of order the curvature scale, and subtracting the zero mode. We regulate this distance function by subtracting the mean:
\be
\delta_{12}=d_{12}-\langle d_{12}\rangle. \label{dist2}
\ee
The convolution will not come into play since we will evaluate all overlaps at late times, where the ratio of horizon size to universe size vanishes. At intermediate times this convolution would come into play.

Recall that the correlator $\langle \hat{\phi}(x_i) \hat{\phi}(y_i)\rangle$ on a constant time slice exhibits an IR divergence, meaning that the vacuum violates cluster decomposition. There is a useful analogy to this in the thermodynamic literature: often times the Boltzmann-Gibbs measure $\mathcal{P}_{BG} \sim e^{-\beta H}$ leads to a violation of cluster decomposition due to a plethora of equilibrium states. The simplest case of which we are aware is the Ising model at low temperatures, where ergodicity is broken due to two ``pure states" of positive and negative magnetization. A pure state is one which satisfies cluster decomposition. In this case, we can perform an explicit decomposition of the Boltzmann-Gibbs measure into the two pure states $\langle \cdots \rangle_+$ and $\langle \cdots \rangle_-$:\footnote{For a discussion of these issues and related ones in the context of spin glasses, see \cite{pedestrians}.}
\be
\langle \cdots \rangle =\frac{1}{2}\langle \cdots \rangle _+ +\frac{1}{2}\langle \cdots \rangle_-\;.
\ee
The distance functions defined above for cosmologies come from analogous definitions in this context, where one can define an overlap between two spin configurations $\sigma_1$ and $\sigma_2$ as 
\be
d_{12}=\frac{1}{N}\sum_{i=1}^N \sigma_1^i \sigma_2^i\;.
\ee
As in the cosmological setting, this is a measure of similarity between two different configurations. Overlaps between different pure states can also be defined. For pure states $\alpha$ and $\beta$ we have 
\begin{align}
d_{\alpha\beta}&=\frac{1}{N}\sum_{i=1}^N \langle \sigma_1^i\rangle_\alpha \langle \sigma_2^i\rangle_\beta \\
&= \frac{1}{Z_\alpha Z_\beta}\int_{\sigma_1\in \alpha}\int_{\sigma_2\in \beta} \mathcal{D}\sigma_1\mathcal{D}\sigma_2\; e^{-\beta H[\sigma_1]}e^{-\beta H[\sigma_2]}\;d_{12}\;,\label{spinglassdist}
\end{align}
where in the second line we have rewritten the overlap between states in terms of the overlap between configurations. This point is \emph{crucial}: one can compute overlaps between states \emph{without} knowing the explicit decomposition into pure states, as is known in the simple case of the Ising model. This is of course useful in the literature of disordered systems, and it will serve useful in our cosmological setting as well. Adapting this analysis to our case simply requires using an appropriate distance definition, which we have given in (\ref{dist1}) and (\ref{dist2}), and replacing the Boltzmann-Gibbs measure with the Hartle-Hawking wavefunctional, since that defines the distribution from which we draw configurations. From this point of view, the violation of cluster decomposition is not so worrisome, as the wavefunctional still gives well-defined overlaps between pure states which do satisfy cluster decomposition. Again, the fact that we cannot perform the decomposition explicitly is irrelevant for these calculations.

In the cosmological setting, we imagine there is some pure state decomposition of $|\Psi[\phi]|^2$ as 
\be
\mathcal{P}[\phi]=|\Psi[\phi]|^2=\sum_{\alpha}w_\alpha \mathcal{P}_\alpha[\phi]\,,
\ee
with the $w_\alpha$ summing to one. In analogy with the spin glass case (\ref{spinglassdist}), we can define a distance between pure states $d_{\alpha\beta}=d[\langle \hat{\phi}\rangle_\alpha, \langle\hat{\phi}\rangle_\beta]$ and rewrite it in terms of distances between configurations. Using the replica trick and the clustering property of the pure states, we can write the probability for finding a renormalized distance $\Delta$ between two states as 
\be
P(\Delta)=\langle \delta(\Delta-\delta_{12})\rangle = \int \mathcal{D}\phi_1 \mathcal{D}\phi_2\, \mathcal{P}[\phi_1]\mathcal{P}[\phi_2]\,\delta(\Delta-\delta_{12})\,.
\ee
A three-point probability $P(\Delta_1,\Delta_2,\Delta_3)$ can be similarly defined. In computing $P(\Delta)$ it is easier to first compute its Laplace transform $G(s)=\langle e^{-s\delta_{12}}\rangle$, and then obtain the original distribution by inverse Laplace transform:
\be
P(\Delta)=\frac{1}{2\pi i}\int_{-i\infty}^{i\infty} e^{s\Delta}G(s)\,ds\;.\label{invlap}
\ee
The calculation of $G(s)$ is the same as in \cite{Anninos:2011kh}, so we just write down the (intermediate) result:
\be
G(s)=\langle e^{-s\delta_{12}}\rangle ={ \prod_{\vec{k}\neq 0}}^{\prime}\f{e^{s/\beta_k}}{1+s/\beta_k}\;.
\ee
The three-point distance distribution is also analogous and becomes 
\begin{align}
G(s_1,s_2,s_3)&=\langle e^{-s_1\delta_{23}}e^{-s_2\delta_{13}}e^{-s_3\delta_{12}}\rangle\\
&={ \prod_{\vec{k}}}^{\prime}\f{e^{(s_1+s_2+s_3)/\beta_k}}{1+(s_1+s_2+s_3)/\beta_k+3(s_1s_2+s_1s_3+s_2s_3)/4\beta_k^2}\;.\label{triple}
\end{align}
It remains to compute these products and perform an inverse Laplace transform to get at the probability distributions. 

As compared to the de Sitter case, we see that we simply have a different $\beta_k$ we need to deal with. At late times we get
\be
\beta_k =\f{L^d}{2} \,k^{d-\theta}\,\ell^{d-\theta-1}\,.
\ee
Since $\theta<0$, the power of $k$ is increased relative to the de Sitter value. This reflects the fact that the infrared modes increasingly dominate as $\theta\rightarrow -\infty$, i.e. as the scale factor $a(t)=t^{q}$, written in terms of non-conformal time, approaches $a(t)=t$. As elucidated in previous work, it is this domination of these infrared modes that leads to Gumbelanity and ultrametricity. Thus, increasing the magnitude of $\theta$ means we should witness an even sharper distribution than in the de Sitter case. We will verify in the next subsections that this is indeed the case.

We would like to stress that the intuition gained by considering formulae in FRW as following from formulae in dS upon the effective shift in dimensionality $d\rightarrow d-\theta$ leads immediately to this result. The substitution should be understood as occurring in Fourier space formulae and not in objects like integrals over momentum space, which of course still occur in $d$ dimensions.

\subsection{FRW$_2$}
We begin by considering 2-dimensional FRW space. Unfortunately, to keep the notation transparent and uncluttered, we defined a transformation to the variable $\theta$ that breaks down when $d=1$, which is the case at hand. Were we to go back and define a different transformation such that our metric became 
\be
ds^2=\eta^{2\theta/d}\left(\f{-d\eta^2+dx_i^2}{\eta^2}\right),
\ee
then for our mode functions we would have 
\be
v_k(\eta)=(-\eta)^\f{d^2+\theta-d\theta}{2d}H^{(1)}_{\f{d^2+\theta-d\theta}{2d}}(k\eta)\,.
\ee
The cluttering of notation is why we avoided this definition of $\theta$. But notice that in this case if we specify to $d=1$, then the $\theta$-dependence drops out and we have $\beta_k=kL/2$, which is just the dS$_2$ value. This is expected since in two dimensions a massless scalar is conformally coupled, and FRW is conformally related to de Sitter space. Thus, the fact that the two-point distance distribution is a Gumbel distribution and the three-point distance distribution exhibits ultrametricity carries over from the calculations in dS$_2$ performed in \cite{Anninos:2011kh}. 

One may wonder about flat space ($\theta=d$), which is also conformally related to these spacetimes (indeed, in two dimensions everything is conformally flat). The same conclusions would apply there as well. The two point function of the massless scalar is logarithmic and does not cluster decompose in any of these spacetimes. 

Two-dimensional FRW would be a simple scenario in which to consider massive fields, which will not be trivially related to the dS$_2$ case considered in \cite{Roberts:2012jw}. However, the wave equation would have to be solved numerically.

Of course, one could also consider conformally coupled fields in dS$_{d+1}$ for $d>1$. As long as the spatial dimension is sufficiently low, these masses are below the bound found in \cite{Roberts:2012jw} and thus exhibit some weak form of ultrametricity. By the conformal equivalence of FRW and dS in arbitrary dimension one can adapt this result to FRW spacetimes with a non-canonical scalar. In the next subsection we will stick to massless scalar fields in higher dimensions. 
\subsection{Higher-dimensional FRW single overlap}\label{triples}
We begin with 
\be
\langle e^{-s\delta_{12}}\rangle ={ \prod_{\vec{k}\neq 0}}^{\prime}\f{e^{s/\beta_k}}{1+s/\beta_k}.
\ee
Approximating the logarithm by an integral, we get 
\be
\log\langle e^{-s\delta_{12}}\rangle \approx \f{1}{2}\int d^d k \f{L^d}{(2\pi)^d}\left[\f{s}{\beta_k}-\log\left(1+\f{s}{\beta_k}\right)\right].\label{approxim}
\ee
We define $\tilde{\beta}_k=k^\theta \beta_k$ to get 
\be
\log\langle e^{-s\delta_{12}}\rangle\approx \f{w_d}{(2\pi)^d\ell^{d-\theta-1}}\int_{\tilde{\beta}_0}^\infty d\tilde{\beta}\left[\f{\tilde{s}}{\tilde{\beta}^{(d-\theta)/d}}-\log\left(1+\f{\tilde{s}}{\tilde{\beta}^{(d-\theta)/d}}\right)\right],\label{mess}
\ee
where
\be
\tilde{s}=2^{\theta/d}L^{-\theta}\ell^{\theta(1-d+\theta)/d}s, \qquad \tilde{\beta}_0=\f{1}{2}(2\pi)^d\ell^{d-\theta-1}.
\ee
Unfortunately, this integral cannot be performed analytically except in special cases. However, we can go back to (\ref{approxim}) and perform the angular integrals, since the integrand only depends on the magnitude of $\vec{k}$. Re-exponentiating and ignoring dimensionless factors, the calculation of $\langle e^{-s\delta_{12}}\rangle$ in arbitrary dimension reduces to an effectively one-dimensional problem:
\be
\langle e^{-s\delta_{12}}\rangle \sim  \prod_{n=1}^\infty\f{e^{s/n^{1-\theta}}}{\left(1+s/n^{d-\theta}\right)^{n^{d-1}}},\label{product}
\ee
where we have allowed ourselves a multiplicative redefinition in $d_{12}$.

We will numerically compute the logarithm of this product by summing a finite number of terms, and then perform the inverse Legendre transform by saddle point approximation. The contour of (\ref{invlap}) is deformed to a steepest descent contour to allow for the saddle point approximation.

\begin{figure}[t!]
\begin{center}
\begin{tabular}{ccccc}
\includegraphics[width=0.4\textwidth]{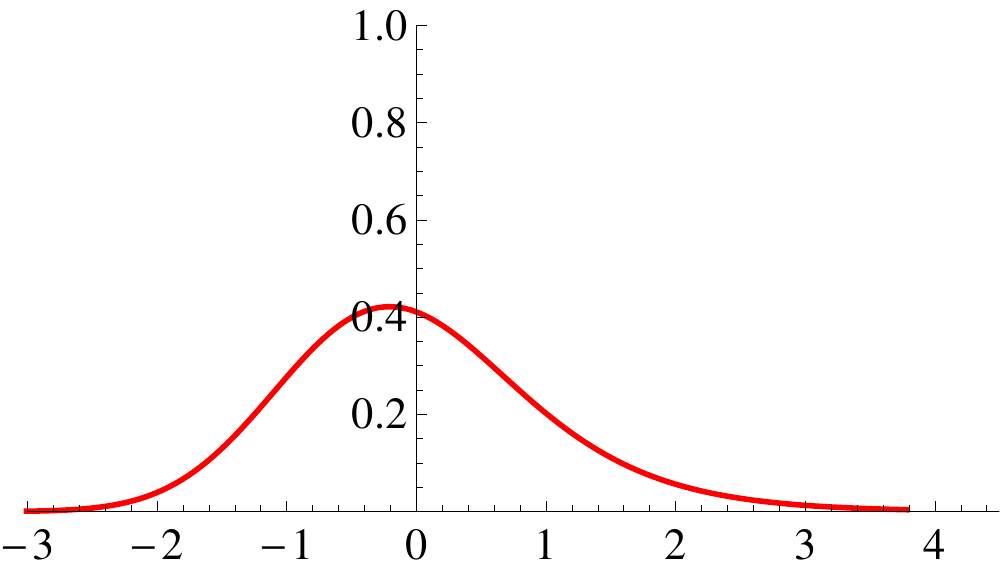} & \phantom{spi} &
\includegraphics[width=0.4\textwidth]{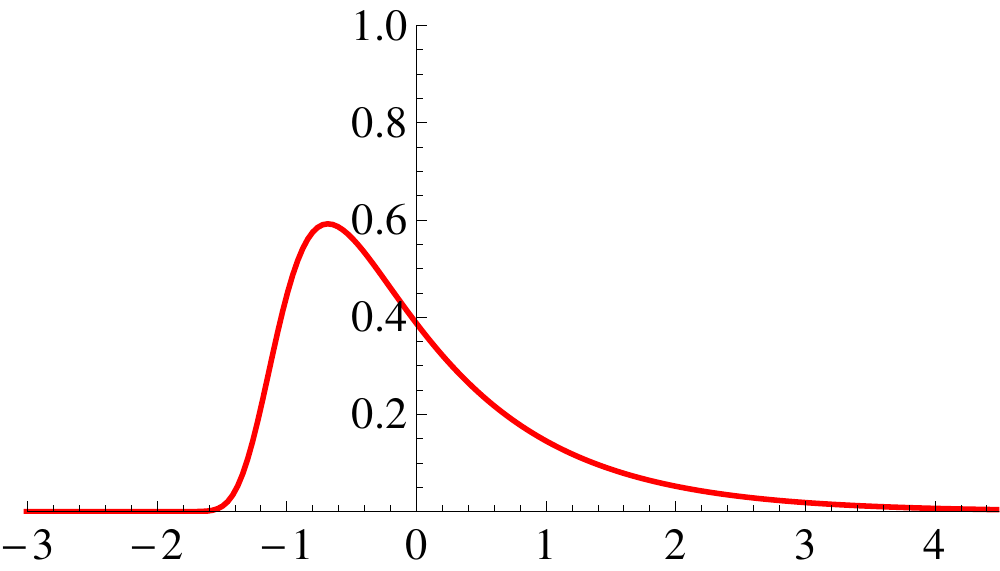} & \phantom{spi} &\\
\\
\includegraphics[width=0.4\textwidth]{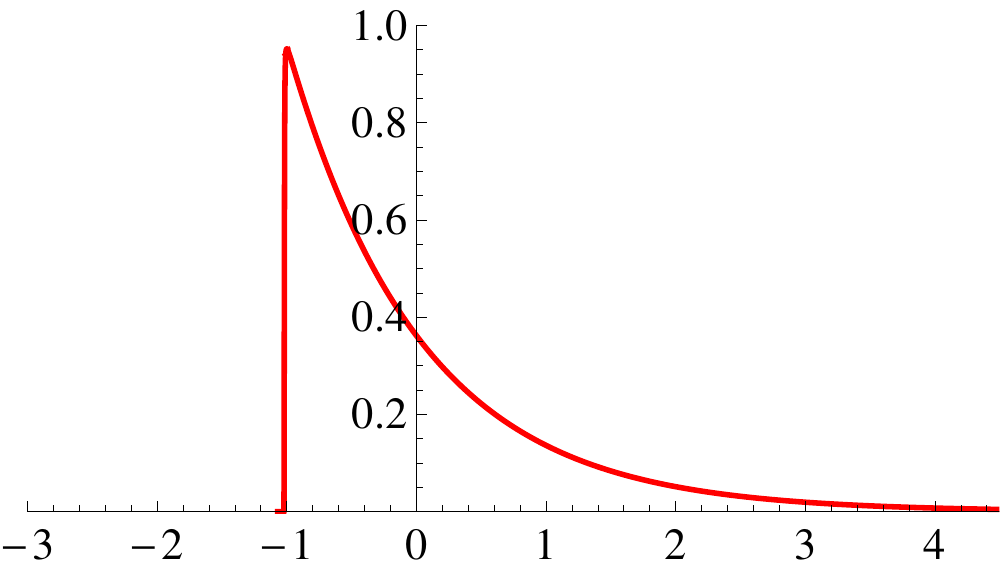}& \phantom{spi} &
\includegraphics[width=0.4\textwidth]{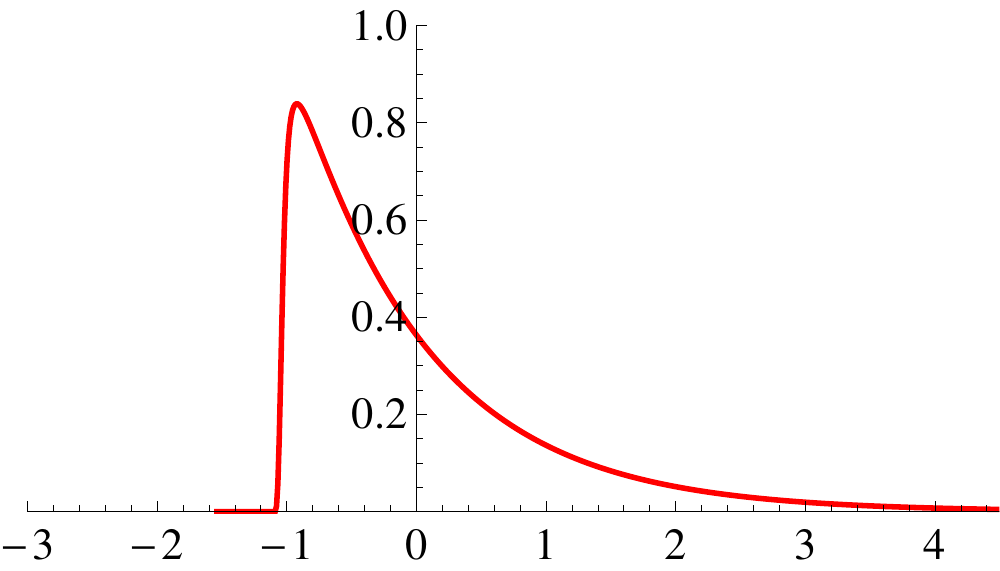} \\
& & & & \\
\end{tabular}
\caption{$P(\Delta)_\theta$ in 3+1 dimensions. Clockwise from the top left, we have $\theta=1.3$, $\theta=0$ (dS), $\theta=-3$, and $\theta=-7$. The case $\theta=1.3$ is not physically relevant for us, but we reproduce it here since it is mathematically the same as a massive scalar field in de Sitter, which begins to lose Gumbelanity, as shown in \cite{Roberts:2012jw}. For $\theta<0$ the curves have semi-infinite support and thus become generalized extreme value distributions (like the Fr\'echet or Weibull distributions). For $\theta\geq 0$ the curves have infinite support.
}\label{gumbels}
\end{center}
\end{figure}

The results are striking: as $\theta$ is made large and negative, the Gumbel distribution of de Sitter approaches a generalized extreme value distribution that exponentially increases until a critical value, at which point it drops straight down. The numerical data for the case of $3+1$ dimensions is presented in Figure \ref{gumbels}.

In fact, as $\theta\rightarrow -\infty$, we can compute $P(\Delta)_\theta$ analytically in any dimension. In this limit, only the first term in the product (\ref{product}) contributes. Thus, we get 
\be
P(\Delta)_{\theta\rightarrow-\infty}=\f{1}{2\pi i}\int_{-i\infty}^{i\infty}G(s)_{\theta\rightarrow-\infty}\;e^{s\Delta}=\f{1}{2\pi i}\int_{-i\infty}^{i\infty}\f{e^{s(1+\Delta)}}{1+s}ds.
\ee
This integral has a simple pole at $s=-1$. For $\Delta<-1$, we can close the contour to the right in the complex $s$-plane. This contour encloses no poles, so the integral vanishes. For $\Delta>-1$, we have to close the contour to the left in the complex $s$-plane. This contour now encloses the $s=-1$ pole, so the integral evaluates to 
\be
P(\Delta)_{\theta\rightarrow-\infty}=e^{-1-\Delta}.
\ee
The full expression becomes this exponential glued onto the constant function $0$ for $\Delta<-1$, i.e. $P(\Delta)_{\theta\rightarrow-\infty}=e^{-1-\Delta}\,\Theta(1+\Delta)$, which is precisely what the numerical plots in Figure \ref{gumbels} seem to be suggesting. This probability distribution integrates to $1$, as required.

This approach can be used to see the asymmetry of the distributions for finite $\theta$ analytically. We have
\be
e^{s\Delta}G(s)_\theta=e^{s\Delta}\prod_{n=1}^\infty\f{e^{s/n^{1-\theta}}}{\left(1+s/n^{d-\theta}\right)^{n^{d-1}}}=\f{e^{s(\zeta[1-\theta]+\Delta)}}{\prod_{n=1}^\infty \left(1+s/n^{d-\theta}\right)^{n^{d-1}}}\label{products}
\ee
One can show that the infinite product in the denominator converges uniformly, by a combination of Raabe's test and the Weierstrass M-test on its sequence of partial sums. Thus, any poles in the resulting function are zeros of one of the product factors of the denominator of the right hand side of (\ref{products}). Similar to the argument for $\theta\rightarrow -\infty$, we now see that $P(\Delta)_\theta$ has no support for $\Delta<-\zeta[1-\theta]$, since we can close the contour to the right and enclose no poles when performing the inverse Legendre transform.  Furthermore, since $\zeta[1-\theta]$ monotonically decreases from infinity to $1$ as $\theta$ goes from $0$ to minus infinity, we see that $P(\Delta)_\theta$ has \emph{less} support as $\theta$ becomes more negative, i.e. as the acceleration decreases. The de Sitter Gumbel has infinite support, and this is reflected here since $\theta=0$ for de Sitter and $\zeta[1]$ diverges. For $\Delta > -\zeta[1-\theta]$ we must close the contour to the left and enclose an infinite number of poles. If $\Delta \gg \big|-\zeta[1-\theta]\,\big|$, evaluating the residues will result in the $e^{s\Delta}$ piece giving the dominant contribution, so the function will be strongly exponentially suppressed since all the poles are at negative $s$. Combining all these observations, we see that $P(\Delta)_\theta$ is generically exponentially suppressed at large and positive $\Delta$, whereas it cuts off at some finite and negative $\Delta$, where the cutoff becomes less negative as the magnitude of $\theta$ is increased. This gives a probability distribution $P(\Delta)_\theta$ whose asymmetry increases as the acceleration decreases. Given that $P(\Delta)_\theta$ has semi-infinite support and a ``light" (i.e. exponential) tail, it is most appropriately identified with a Weibull distribution.

\subsection{Higher-dimensional FRW triple overlap}
\begin{figure}
\begin{center}
{ \includegraphics[scale=.77]{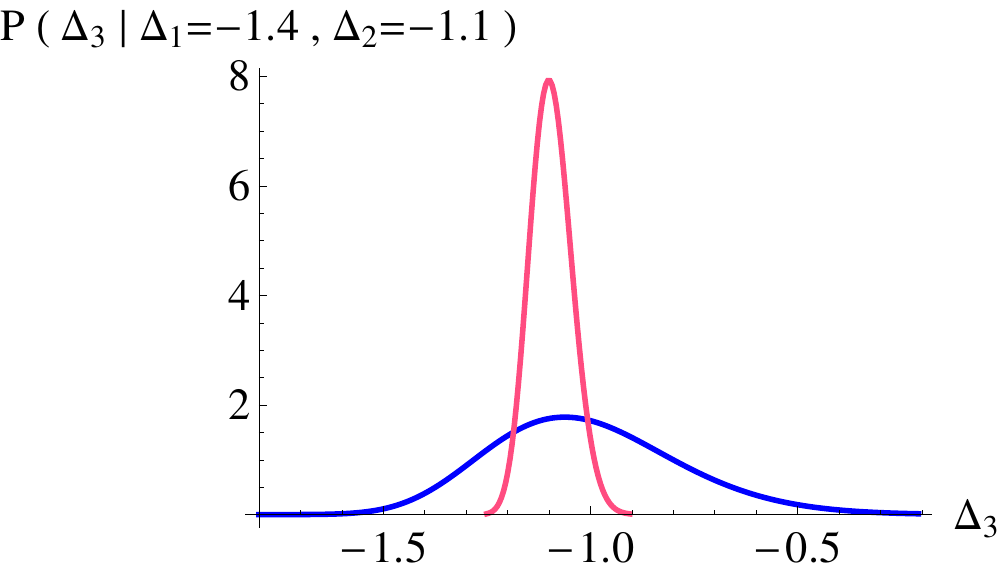} \includegraphics[scale=.77]{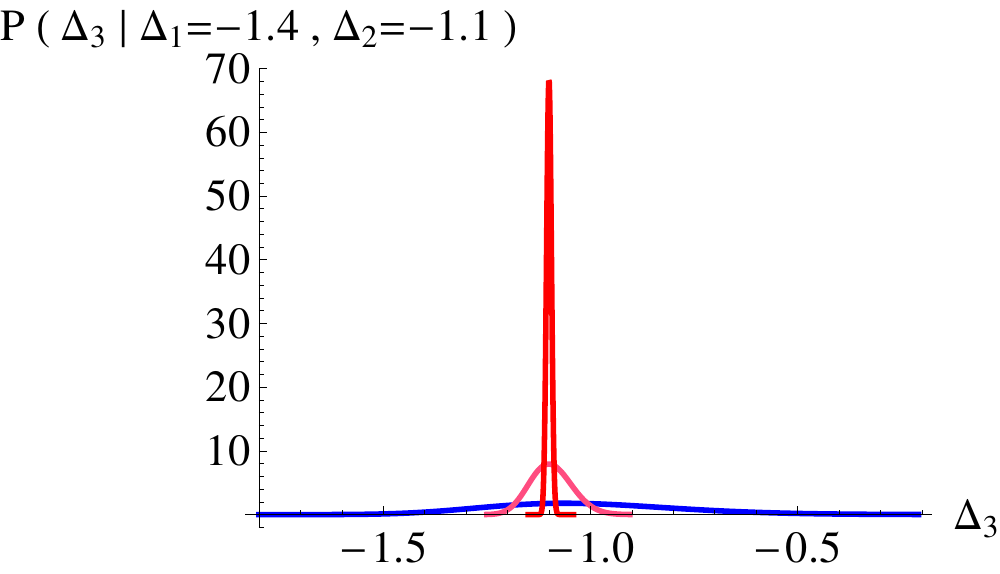}\vspace{5mm} \includegraphics[scale=.77]{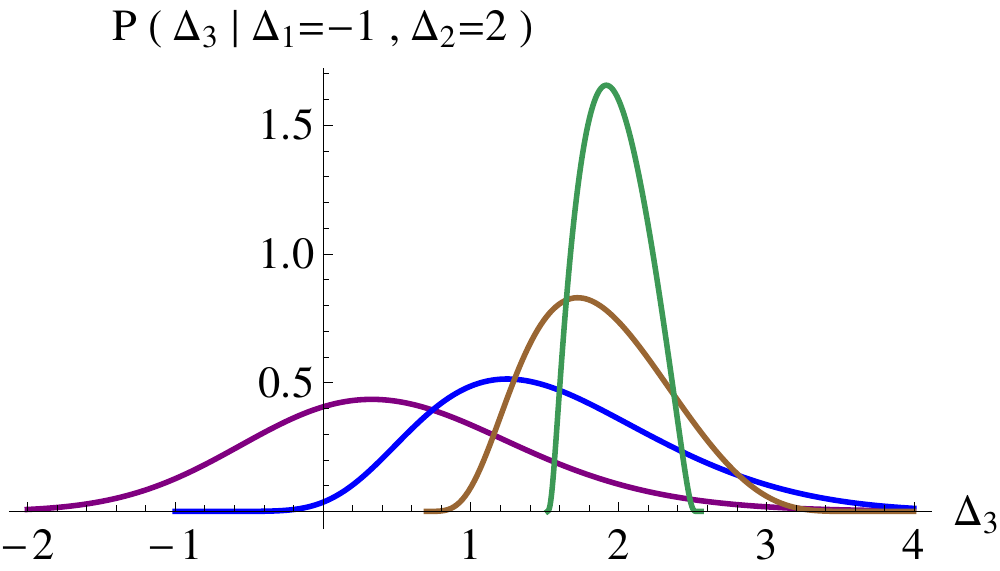}}
\caption{{\bf Top left:} Here we pick $\Delta_1=-1.4$ and $\Delta_2=-1.1$. The blue curve is for $\theta=0$ and the pink curve is for $\theta=-1$. The FRW case is more indicative of ultrametricity than the de Sitter case since it peaks more sharply on max$(-1.4,-1.1)=-1.1$. However, according to the analytic arguments in the main text, we cannot take $\theta$ arbitrarily negative since $\Delta_1<-1$ and $\Delta_2<-1$ . For the $\theta=-1$ case exhibited here we have $\zeta[1-\theta]=\zeta[2]=\pi^2/6\sim 1.6$ so we have some breathing room until we hit $\Delta_i = -\zeta[1-\theta]$, which is the point at which the conditional problem we have set up breaks down due to conditioning on an impossibility. {\bf Top right:} Same plot as top left, except we have added the red curve which corresponds to $\theta=-1.35$, giving $-\zeta[1-\theta]\approx -1.41$, which is very close to min$(-1.4,-1.1)=-1.4$, so $P(-1.4)$ is incredibly suppressed. {\bf Bottom:} Here we display a conditional probability upon picking $\Delta_1=-1$ and $\Delta_2=2$. The purple curve is for $\theta=1.5$, the blue curve is for $\theta=0$, the brown curve is for $\theta=-2$, and the green curve is for $\theta=-7$. Notice that as $\theta$ decreases the conditional proabability peaks on max$(-1,2)=2$, as required for ultrametricity. For these values of $\Delta_1$ and $\Delta_2$ the de Sitter case shows little evidence of ultrametricity, whereas the probability peaks more and more sharply as the acceleration slows. The $\theta=1.5$ case is unphysical in our context but is included as a check, since it is equivalent to a massive field in de Sitter, to see that it shows even less evidence for ultrametricity than the de Sitter case.}\label{ultrametriccombos}
\end{center}
\end{figure}

We would now like to compute triple overlaps to study to what extent the ultrametric structure found in de Sitter is preserved in accelerated FRW. Performing the angular integrals as before, we get 
\be
G(s_1,s_2,s_3)_\theta\approx\prod_{n=1}^{\infty}\f{e^{(s_1+s_2+s_3)/n^{1-\theta}}}{\left(1+(s_1+s_2+s_3)/n^{d-\theta}+3(s_1s_2+s_1s_3+s_2s_3)/4n^{2(d-\theta)}\right)^{n^{d-1}}}
\ee
in $d+1$ dimensions. We can calculate this as before by taking a finite number of the terms and performing the inverse Laplace transforms by saddle point approximation. Examples that illustrate the ability of accelerated FRW cosmologies to accentuate the structure seen in the de Sitter case are presented in Figure \ref{ultrametriccombos}. Understanding the sharper peaking in e.g. the right panel can be understood by analogy to the de Sitter case: there, peaks sharpen as $\Delta_1$ and/or $\Delta_2$ are made more negative, i.e. more suppressed regions of parameter space corresponding to similar configurations are sampled. Recall that similar (dissimilar) configurations correspond to negative (positive) values of the overlap. Alternatively, one can mimic sampling these rare, similar configurations by keeping $\Delta_1$ and $\Delta_2$ fixed and taking $\theta$ large and negative. To see this in a simple example, fix $\Delta=-1.4$ and observe $P(\Delta)_\theta$ as $\theta$ is made more negative in Figure \ref{gumbels}. 

\begin{figure}
\begin{center}
{\includegraphics[scale=.77]{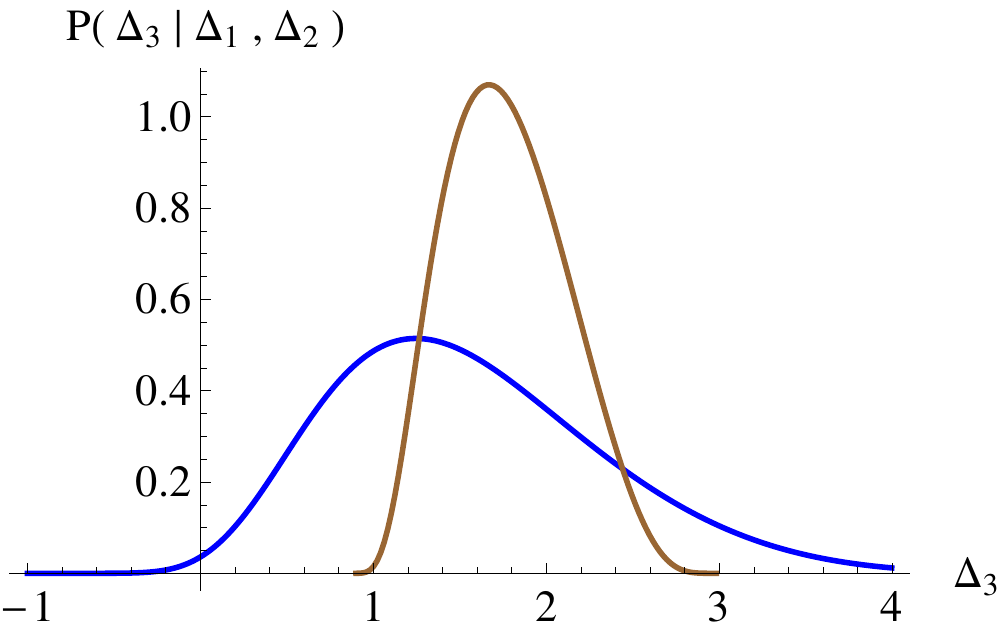}  \includegraphics[scale=.77]{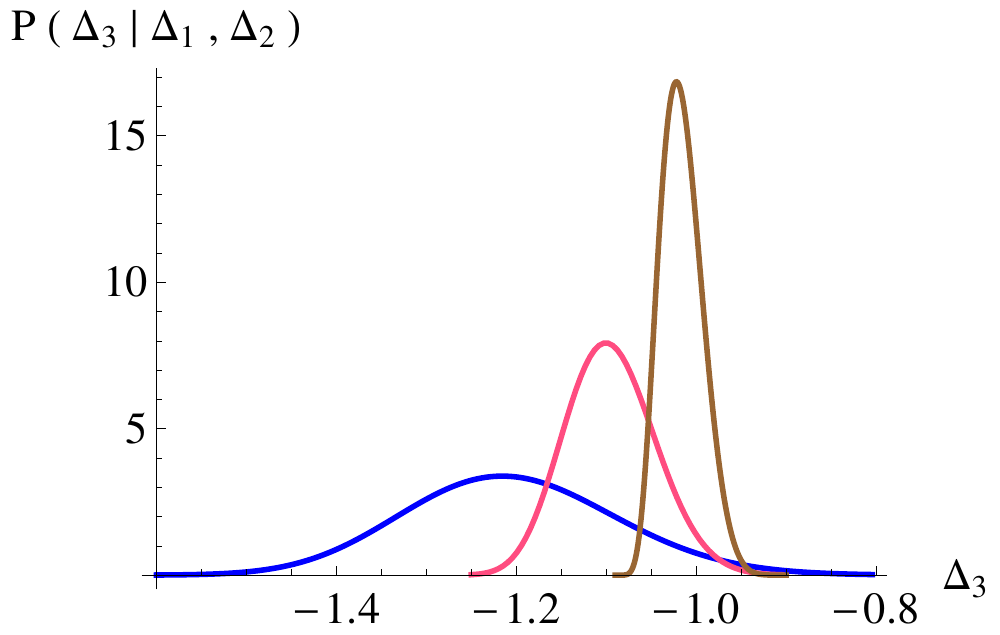}}
\caption{{\bf Left:} Here we display a conditional probability upon picking $\Delta_1=-1$ and $\Delta_2=2$ for $\theta=0$, given by the blue curve, and $\Delta_1= -1$ and $\Delta_2= 1.84$ for $\theta=-3$, given by the brown curve. These values satisfy $P(\Delta_i, \theta=0)\approx P(\Delta_i, \theta=-3)$. Like in Figure \ref{ultrametriccombos}, the FRW case shows more evidence of ultrametricity than the de Sitter case.
{\bf Right:} Here we pick $\Delta_1=-2.12$ and $\Delta_2=-1.22$ for $\theta=0$, given by the blue curve, $\Delta_1=-1.4$ and $\Delta_2=-1.1$ for $\theta=-1$, given by the pink curve, and $\Delta_1=-1.07$ and $\Delta_2=-1.02$ for $\theta=-3$, given by the brown curve. Again, these values satisfy $P(\Delta_i, \theta=0)\approx P(\Delta_i, \theta=-1)\approx P(\Delta_i, \theta=-3)$ and the FRW case shows more evidence of ultrametricity than the de Sitter case.}\label{probfixed}
\end{center}
\end{figure}

The observation of the previous paragraph suggests that to make a meaningful comparison between de Sitter and FRW when talking about sharpness of ultrametric structure, one needs to weight the ultrametric structure by how rare a region of parameter space is being sampled. Thus, instead of fixing $\Delta_1$ and $\Delta_2$ we will instead pick a $\Delta_1$ and $\Delta_2$ for some $\theta$, and as we vary $\theta$ we will keep $P(\Delta_1)_\theta$ and $P(\Delta_2)_\theta$ fixed. Recall from Figure \ref{gumbels} that the the function $P(\Delta)_\theta$ for any $\theta$ is two-to-one, but since the function $P(\Delta,\theta)\equiv P(\Delta)_\theta$ is continuous in $\theta$ we simply pick the distances $\Delta_1$ and $\Delta_2$ which are continuously connected to the original ones as we vary $\theta$. Setting the problem up this way only attenuates the degree to which FRW improves upon de Sitter; as seen in Figure \ref{probfixed} the peaks for FRW remain both sharper (i.e. more ``precise") and more correctly localized on max$(\Delta_1,\Delta_2)$ (i.e. more ``accurate").

Similar to the single overlap, for $\theta\rightarrow -\infty$ in arbitrary dimension we can make some analytic remarks. In this case only the first term in the product contributes and we have
\be
P(\Delta_1,\Delta_2,\Delta_3)_{\theta\rightarrow -\infty}\approx\int_{\gamma_1-i\infty}^{\gamma_1+i\infty}ds_1\int_{\gamma_2-i\infty}^{\gamma_2+i\infty}ds_2\int_{\gamma_3-i\infty}^{\gamma_3+i\infty}ds_3\;\f{e^{s_1(1+\Delta_1)+s_2(1+\Delta_2)+s_3(1+\Delta_3))}}{1+s_1+s_2+s_3+3(s_1s_2+s_1s_3+s_2s_3)/4}.
\ee
We see immediately that at least some of the ultrametric structure found in de Sitter space vanishes in this limit: if $\Delta_1$, $\Delta_2$, and $\Delta_3$ are all less than $-1$, then we can set $\gamma_i=0$ and close all three contours to the right, i.e. choose the 3-cycle in $\mathbb{C}^3$ that encloses the ``positive octant" with Re$(s_i)>0$. This 3-cycle enclose no poles, since we need to have Re$(s_i)<0$ for at least one of the $s_i$ for the denominator to vanish. Thus, evaluating something like $P(\Delta_3|-1.4,-1.1)_{\theta\rightarrow-\infty}$, we see that not only does it \emph{not} peak on max$(-1.4,-1.1)=-1.1$, as would be indicative of ultrametricity, but it has no support for $\Delta_3<-1$. This is not surprising since in this limit the question is both ill-defined and trivial. By requiring $\Delta_1$ and $\Delta_2$ to be less than $-1$, one is conditioning on an impossibility, as $P(\Delta)_{-\infty}$ has no support for $\Delta<-1$. Even if the conditioning were well-defined, we already know that the single overlap $P(\Delta)_{\theta\rightarrow-\infty}$ has no support for $\Delta<-1$, so a conditional probability will not change that result. These observations mean we should impose a general restriction, which we will state at the end of the next paragraph.

For finite $\theta$ we can still make some analytic arguments analogous to the single overlap:
\begin{align}
G(s_1,s_2,s_3)_\theta \;e^{\sum_i s_i\Delta_i}&=\f{e^{s_1\Delta_1+s_2\Delta_2+s_3\Delta_3}\prod_{n=1}^{\infty}e^{(s_1+s_2+s_3)/n^{1-\theta}}}{\prod_{n=1}^{\infty}\left(1+(s_1+s_2+s_3)/n^{d-\theta}+3(s_1s_2+s_1s_3+s_2s_3)/4n^{2(d-\theta)}\right)^{n^{d-1}}}\nonumber\\\nonumber \\
&=\f{e^{s_1(\zeta[1-\theta]+\Delta_1)+s_2(\zeta[1-\theta]+\Delta_2)+s_3(\zeta[1-\theta]+\Delta_3)}}{\prod_{n=1}^{\infty}\left(1+(s_1+s_2+s_3)/n^{d-\theta}+3(s_1s_2+s_1s_3+s_2s_3)/4n^{2(d-\theta)}\right)^{n^{d-1}}}.\label{tripleprod}
\end{align}
For $\Delta_i<-\zeta[1-\theta]$, in the calculation of $P(\Delta_1,\Delta_2,\Delta_3)_\theta$ we can pick the same 3-cycle as above and enclose no poles, thus giving no support. The top right panel of Figure \ref{ultrametriccombos} shows, however, that for an illustrative example the ultrametric peak does sharpen at finite $\theta$ until this point is reached. Again, the vanishing of $P(\Delta_1,\Delta_2,\Delta_3)_\theta$ in this limit is not surprising since the values of $\Delta_i$ are such that $P(\Delta_i)_\theta=0$. This means we need to impose a restriction on the conditional probability we are setting up. When computing $P(\Delta_3|\Delta_1, \Delta_2)_\theta$, we should ensure that $\Delta_1$ and $\Delta_2$ are both greater than $-\zeta[1-\theta]$. If not, we would be conditioning on something that is impossible, according to the arguments of the previous section. Upon this conditioning, it is well-defined to compute $P(\Delta_3|\Delta_1,\Delta_2)_\theta$ for a $\Delta_3$ which is less than $-\zeta[1-\theta]$, but the answer will always vanish.

\begin{figure}
\begin{center}
{\includegraphics[scale=1]{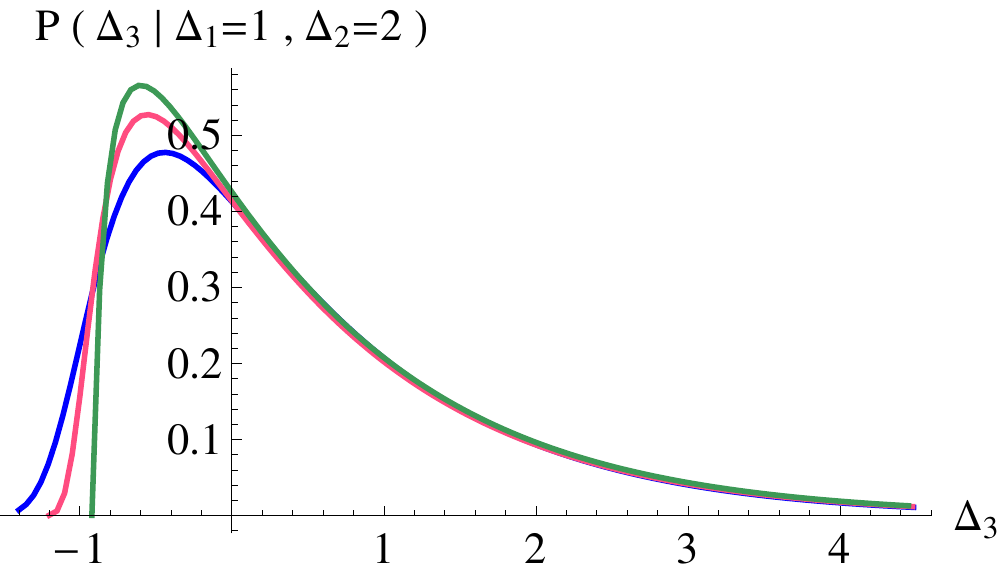}}
\caption{Here we display a conditional probability upon picking $\Delta_1=1$ and $\Delta_2=2$. The blue curve is for $\theta=0$, the pink curve is for $\theta=-1$, and the green curve is for $\theta=-7$. The peak sharpens as usual as $\theta$ gets negative, but none of these cases shows evidence of ultrametricity.}\label{ultrametriccombopos}
\end{center}
\end{figure}

The power of FRW to accentuate the ultrametric structure seen in the de Sitter case is not unlimited. If we sample points to the right of the peak of the generalized extreme value distributions in Figure \ref{gumbels}, e.g. $\Delta_1=1$ and $\Delta_2=2$, for which the de Sitter case shows virtually no evidence of ultrametricity, then taking $\theta$ negative does not help the state of affairs, as exhibited in Figure \ref{ultrametriccombopos}. This is rooted in the fact that the single overlap $P(\Delta)_\theta$ does not show strong dependence on $\theta$ for $\Delta>0$. Thus, the restriction to sampling relatively similar configurations to see an ultrametric structure emerge remains.

Finally, a curious property to note is that the results for these overlaps have a symmetry transform under $\beta_k\rightarrow -\beta_k$ and $\Delta_i\rightarrow -\Delta_i$. In other words, $P(\Delta;\beta_k)=P(-\Delta;-\beta_k)$ and $P(\Delta_1|\Delta_2,\Delta_3; \beta_k)=P(-\Delta_1|-\Delta_2,-\Delta_3; -\beta_k)$. This leads to results which give an inverse ultrametric structure, meaning all triangles are isosceles with the unequal side the \emph{longest} of the three. In this case, dissimilar configurations are most suppressed and it is conditioning on sampling such configurations which leads to the inverse ultrametric structure. It is unclear if such a structure comes from any interesting physical scenario, although such a $\beta_k$ is obtained by e.g. analytic continuation from dS$_4$ to EAdS$_4$, where one treats the direction perpendicular to the boundary as Euclidean time. It should be stressed that Lorentzian AdS and hyperscaling-violating spacetimes do not have such a dynamical structure in time, but possibly a suitably modified version of the above calculations, where time is replaced by the holographic radial coordinate, will show such a structure emerge.

\begin{section}
{$\theta\rightarrow-\infty$ and its connection to little string theory}\label{infinity}
\end{section}
\noindent Given the interesting structure of the $\theta\rightarrow -\infty$ FRW geometry in the context of single and triple field overlaps, we shall try to analyze it in a little further detail. This is just a curvature-dominated FRW cosmology with linear scale factor, which has been extensively studied, so we will focus instead on its analytic continuation to a hyperscaling-violating geometry. One can imagine/hope that the structure uncovered in the previous section can also be understood as existing in some form in AdS and hyperscaling-violating geometries, and it is with this view in mind that we study the hyperscaling-violating geometries.

For hyperscaling-violating geometries, the double scaling limit $\theta\rightarrow -\infty$ and $z\rightarrow \infty$ with $-\theta/z=\eta>0$ fixed has been studied to some extent due to having a spectral density that is not exponentially suppressed \cite{Hartnoll:2012wm}. The spacetime is conformal to AdS$_2\times \mathbb{R}^2$ and appears in near-horizon geometries of dilatonic black holes \cite{Gubser:2009qt}, in addition to having other interesting properties \cite{Gouteraux:2011ce, Anantua:2012nj}. However, in our case we would like to keep $z$ finite and take $\theta\rightarrow -\infty$, which does not seem to have been studied. In this limit the effect of the dynamical critical exponent $z$ washes out and Lorentz invariance is restored. It is also equivalent to taking $\eta\rightarrow \infty$. The contribution of the Maxwell field drops out of the Einstein-Maxwell-dilaton theories which produce hyperscaling-violating geometries, and the on-shell action limits smoothly to a self-interacting scalar minimally coupled to gravity. Starting from the metric in the form (\ref{hyp1}), we end with the null-energy-condition-satisfying spacetime
\be
ds^2_{d+2}=d\tilde{r}^2+\frac{\tilde{r}^2}{\ell^2}\left(-dt^2+dx_i^2\right)=e^{2\hat{r}/\ell}\left(-dt^2+d\hat{r}^2+dx_i^2\right)\label{infgeom},
\ee
where we have reinstated the curvature scale. Taking $\theta\rightarrow +\infty$ gives the same result. The scale transformation in this limit corresponds to a simple shift in the radial coordinate $\hat{r}$ ($\hat{r}\rightarrow \hat{r}+\lambda$) or a multiplicative rescaling of $\tilde{r}$ ($\tilde{r}\rightarrow \lambda \tilde{r}$). Neither time nor space scales and the effective dimensionality of a hypothetical dual is infinite (we imagine taking $\theta\rightarrow -\infty$ instead of $\theta\rightarrow +\infty$, and thus refer to the effective dimensionality as large or infinite, since this is the direction in which the ultrametric structure of the FRW cosmologies sharpens). It would be interesting to see how this growth in effective dimensionality is connected to the fact that ultrametricity generically sharpens as the dimension grows \cite{244332}. The higher curvature corrections of this special geometry behave similarly to the case of finite $\theta$ and are presented in Appendix \ref{proof2}. The limit $\theta\rightarrow -\infty$ is also strange from the point of view of the entropy density, which in general behaves as
\be
s\sim T^{\frac{d-\theta}{z}}.
\ee
It seems the entropy density, and thus the specific heat, is diverging in this limit. Indeed, for the metric ansatz (\ref{infgeom}) we can solve for black brane geometries, which take the form 
\be
ds^2=\f{dr^2}{r^4\left(1-\;\f{r^2}{r_h^2}\right)}+\frac{1}{r^2}\left(-\left(1-\; \f{r^2}{r_h^2}\right)dt^2+dx_i^2\right).
\ee
Here we are using yet a third radial coordinate $r=1/\tilde{r}$, for which the horizon occurs at $r_h$. Curiously, the emblackening factor is the same as for the static patch of de Sitter. For $r_h=\infty$, i.e. no black brane, the spacetime has a curvature singularity as $r\rightarrow \infty$. By the standard arguments, one can see that the temperature of this black brane is non-vanishing and $r_h$-independent. In other words, changing the horizon location changes the entropy density but does not change the temperature! Mathematically, we can understand this by the scaling properties of the metric: taking $r\rightarrow \lambda r$ leads to an overall factor on the metric and an effective shift in $r_h$. Since overall factors do not affect the temperature, we now understand this strange feature.\footnote{I am indebted to Xi Dong and Gonzalo Torroba for discussions about this spacetime and its cosmological cousin.} The Euclidean continuation of the background with no black brane also has a conical deficit, which is in line with the above reasoning.

In fact, this background is nothing more than the gravity background dual to little string theory, a nonlocal, non-gravitational theory presumably described by strings.\footnote{See \cite{Kutasov:2001uf} for an introductory review of little string theory.} Notice that due to the constant temperature one can integrate up the equation of state $dE=TdS$ to obtain $E=TS$. This gives a Hagedorn density of states $w(E)\sim e^{E\beta}$, which is the same density of states as for a hot gas of free strings. This background can be obtained in the decoupling limit of a stack of NS$5$-branes in Type IIB string theory (it is important to note that $\alpha'$ does not vanish in the decoupling limit as it usually does when decoupling D-branes), where the six-dimensional worldvolume theory is the little string theory. In the $S$-dual picture valid in an intermediate energy range we have a stack of D$5$ branes. In the IR the IIB theory flows to six-dimensional weakly coupled super Yang-Mills with $(1,1)$ supersymmetry and gauge group $U(N)$. When constructed in string theory and compactified on tori, the non-gravitational duals enjoy a T-duality symmetry, which is a further reason why the duals are believed to be nonlocal and described by strings.

The entanglement entropy properties of the dual to this spacetime are the same as for the ones conformal to AdS$_2\times \mathbb{R}^2$. Although the latter incorporates $z$, for time-independent scenarios $g_{tt}$ does not change the results since one is to work on a constant time slice. This explains the curious connection pointed out in \cite{Hartnoll:2012wm} between the entangling properties of geometries conformal to AdS$_2\times \mathbb{R}^2$ and the results of \cite{Ryu:2006ef}, which computed minimal surfaces for the NS$5$-brane background. It is known that a slab geometry on the boundary exhibits a transition from connected to disconnected minimal surfaces whereas a disc geometry does not. This is a strange intermediate behavior between confined and deconfined phases, which is precisely where a Hagedorn spectrum takes over in confining gauge theories. Since the dual theory is expected to have non-gravitational strings, it would be curious if this background were useful in describing the proliferation of gauge theory flux tubes. Here we are imagining an effective description where T-duality is not a symmetry and the extensivity of the entanglement entropy (shown at the end of this section) is not interpreted as a fundamental nonlocality.

It has been shown in \cite{Kulaxizi:2012gy} that upon including a Gauss-Bonnet term in the bulk, connected minimal surfaces exist for both disc and slab geometries, indicating a deconfined phase. It would be interesting to see if the behavior of black brane geometries in the Gauss--Bonnet-corrected bulk theory correlate with this behavior by giving a non-Hagedorn density of states.

The nonlocal nature of the dual can be established without resorting to the brane construction, by considering the UV scaling of the entanglement entropy. The volume law indicative of nonlocality was already shown in \cite{Barbon:2008ut} by using trial surfaces, but one need not resort to trial surfaces (which can sometimes give incorrect leading order behavior) to see the volume law. Consider a time slice of the above background with the curvature scale set to $1$, which has induced metric
\be
ds^2=dr^2+r^2(dy^2+dx_i^2)=dr^2+r^2(d\rho^2+\rho^2d\Omega_{d-1}^2).
\ee
Renaming $\rho\rightarrow \theta$ and performing an identification $\theta\sim\theta+2\pi$ resolves the conical deficit near $r=0$ and gives a metric that can be approximated by flat space for small $\theta$:
\be
ds^2=dr^2+r^2(d\theta^2+\theta^2d\Omega_{d-1}^2)\approx dr^2+r^2(d\theta^2+\sin^2\theta \;d\Omega_{d-1}^2) , \qquad \theta\ll 1.
\ee
The UV boundary lives at large $r$, which is a $d$-sphere as long as we work at small $\theta$. The topological identification is immaterial in this regime. Thus, if our entangling surface is a patch of the sphere localized at small $\theta$, we should recover volume-law scaling due to the lack of warping. To see the details of such a flat-space calculation, see \cite{Li:2010dr}. The similarity of this spacetime to flat space is connected to the nonlocality of the dual theory, although the geometry can also be interpreted as describing the IR phase of a fundamentally local theory.\\

\begin{section}
{Summary and outlook}\label{conclusions}
\end{section}
\noindent The primary purpose of this paper has been to emphasize thinking about FRW cosmologies as an analytic continuation of hyperscaling-violating spacetimes and utilizing the scale covariance of both. We exhibited two distinct scenarios where this point of view proved fruitful: first, in the analysis of higher curvature corrections, and second, in the study of wavefunctionals, correlation functions, holography, and state space overlaps of the sort pioneered by Anninos and Denef. It is clear that the nonperturbative quantum gravitational dynamics in these two backgrounds are two different beasts altogether, but for some more modest perturbative dynamics and analysis a unified approach may be a useful guiding principle.

More concretely, we have shown that scale-covariant geometries of the type analyzed in this paper have a simple structure for the form of higher curvature corrections. We argued that if there exists an action that produces one of these geometries as a solution, then adding a higher curvature term to that action will lead to equations of motion that cannot be solved with a scale-covariant metric ansatz. As a practical matter, in the realm of perturbatively small higher curvature corrections, one simply picks up perturbatively small corrections that break the type of scale covariance we have been analyzing. For example, it is possible that, e.g. in the FRW case, one maintains scale covariance of the form $a(\eta)=\eta + \epsilon \eta^3+\cdots$. But the scale covariance we imagine breaking is the pure power law kind. For the hyperscaling-violating metrics, one can furthermore imagine that the scale covariance is broken to begin with by assuming the full geometry is asymptotically AdS in the UV, or by heating up the system with a black hole. In this context there will be regimes where approximate scale covariance exists (deep in the IR for the asymptotically AdS case and far from the black hole in the finite temperature case), and the perturbative corrections are not physically relevant since the conformal isometry is not exact to begin with.

In the context of the hyperscaling-violating geometries produced from Einstein-Maxwell-dilaton actions, there is a more important point to be made: $\alpha'$ corrections are becoming important for the electric solutions, which means that one cannot dismiss the higher curvature terms as a perturbative breaking of the scale covariance since their contributions become large. And once any higher order corrections become important, then one must consider all of them. It is in this context that the arguments of this paper provide the most stringent constraints. In the absence of loopholes as discussed in Section \ref{loop} and other parts of the main text, this seems to provide further evidence that this geometry generically cannot provide a stable phase in the IR.

In the context of the FRW geometries, the arguments are more of a technical point. In the regimes of e.g. matter or radiation domination there is no reason to expect the higher curvature corrections to become large. 

We then turned to the structure of the state space in accelerated FRW cosmologies, where we accepted the Bunch-Davies vacuum as a sensible starting point for constructing field overlaps. The technology used was directly imported from the literature of disordered systems and massaged to suit the current context. The substitution $d_{\textrm{eff}}=d-\theta$, often used in the case of hyperscaling-violating geometries, proved useful in predicting the structure of these overlaps. The single overlap, which produced a Gumbel distribution in the case of de Sitter, became a generalized extreme value distribution similar to a Weibull distribution and limited to a totally asymmetric distribution (a special case of the Weibull distribution) as the scale factor became linear, i.e. $\theta\rightarrow -\infty$. Furthermore, the triple overlap exhibits even sharper ultrametric structure than de Sitter as the acceleration slows. The numerics indicate that as long as at least one of the $\Delta_i$, say  $\Delta_1$, is sampled to the left of the peak of the generalized extreme value distribution, then taking $\theta$ close to the solution of $-\zeta[1-\theta]=\Delta_1$ will lead to incredibly sharp peaking, approaching a delta function peaked on the value necessary to preserve the ultrametric inequality. See the top right panel of Figure \ref{ultrametriccombos}. It is interesting to note that glassy systems have been modeled with a finite $\theta$ in the past \cite{Fisher:1986zzb}, although this may be a red herring for many reasons, one of which is that such models do not contain the ultrametric structure seen in the Parisi analysis of mean-field spin glasses.

Stated more generally, we argued in Section \ref{triples} that the FRW enhancement of the de Sitter ultrametric structure occurs as long as increasing the magnitude of $\theta$ leads to sampling more suppressed regions of parameter space and the conditional probability remains well-defined. This is in agreement with how one enhances the ultrametric structure at $\theta=0$ (or any fixed $\theta$), which is accomplished by taking the conditioned distance $\Delta_1$ and/or $\Delta_2$ large and negative. This is sampling relatively similar configurations, whereas positive $\Delta_i$ would be sampling relatively dissimilar ones. In the case of the de Sitter Gumbel we see that sampling similar configurations leads to a super-exponentially suppressed region of parameter space, which is where the ultrametric structure is sharpest. It is as if $\theta$ is just acting to renormalize the distances $\Delta_i$, although (\ref{tripleprod}) shows that this is only a useful heuristic. Plotting the conditional probabilities while keeping $P(\Delta_1)_\theta$ and $P(\Delta_2)_\theta$ fixed seems a sensible way to compare different values of $\theta$, and we saw in Figure \ref{probfixed} that dialing $\theta$ still led to an enhancement of ultrametricity. 

It is not yet well understood why one should have to sample relatively similar configurations to see such an ultrametric structure emerge. We seem to have enough data to begin extracting some useful physical lessons from these results, such as how the bulk ultrametric structure is encoded in the boundary theory, but there are indeed more calculations which can be done. An obvious generalization of these calculations is to massive fields. A more interesting problem is to understand the implications for the graviton, which in transverse traceless gauge is effectively a massless scalar. It would be very interesting if these calculations shed light on the nature of the spacetime geometry at $\mathcal{I}^+$. Finally, a less obvious generalization is to analyze anisotropic cosmologies with an aim toward computing field overlaps in Kasner spacetimes, which describe the classical background near the BKL singularity \cite{Belinsky:1970ew}. It was conjectured that a backreacted massless scalar can lead to non-oscillatory behavior near the BKL singularity that is easier to analyze \cite{Belinsky:1973:me}. In any case, the spatial dynamics is expected to decouple point by point near the singularity and lead to ``ultralocality." Since the conjecture, the rich mathematical structure of the classical dynamics has been analyzed, invoking Coxeter groups and Kac-Moody algebras,\footnote{See \cite{Henneaux:2007ej} for a nice review of the classical dynamics and mathematics behind BKL singularities.} but little of the quantum dynamics has been understood outside of loop quantum cosmology \cite{Ashtekar:2009vc}. Perturbative wavefunctionals and field overlaps may be a useful way to begin such a quantum analysis.

We also commented on the possibility of realizing holographic descriptions of FRW spacetimes, either on their own or as part of a cosmological RG flow. Covariant entropy bounds suggest the possibility of holographically describing an accelerated FRW spacetime with a dual theory at $\mathcal{I}^+$. This Q-space/QFT correspondence is proposed to be given by $\Psi_{HH}=Z_{QFT}$, with the dual theory violating hyperscaling. The argument for hyperscaling violation extends to decelerated phases and implies that matter domination and radiation domination in our universe correspond to phases of a dual RG flow which violate hyperscaling. This picture, if correct, helps fill in Strominger's vision of the cosmological evolution of the universe as inverse RG flow \cite{Strominger:2001gp}. Much remains to be understood in this Q-space/QFT correspondence, such as the behavior of higher spin fields and holographic renormalization. An enlightening approach to the latter would be to perturb around the fixed point $\theta=0$, analogous to what was done for e.g. Lifshitz backgrounds in \cite{Korovin:2013bua}. Another concrete calculation would be to study the asymptotic symmetry group of the spacetime, which to the author's knowledge has not been done either for Q-space or hyperscaling-violating spacetimes. As in the de Sitter case, it remains unclear how the bulk ultrametric structure is encoded in the dual theory at any given $\h$. However, the parameter $\theta$ is a new knob one can turn to sharpen the ultrametricity in the bulk and try to understand how a holographic dual should respond.

Finally, given the motivation for understanding geometries with large and negative $\theta$ due to their interesting properties in the case of cosmological field overlaps, we studied the analytic continuation of the cosmology with $\theta=-\infty$, i.e. linear scale factor. In the hyperscaling-violating family the dual to this geometry has an infinite effective dimensionality, and the geometry itself has been encountered before as the gravitational dual to little string theory, the purported worldvolume theory on a stack of NS$5$-branes. The limit $\theta\rightarrow - \infty$ ($d_{\textrm{eff}}\rightarrow \infty$) may be a useful way to think about how nonlocal or string-like dynamics may emerge from particle-like dynamics. Indeed, one can hope that the numerical coefficient of the Brown-York stress tensor in (\ref{brownyork}), which we have suppressed but is given by $d-1-\h$, does not get rescaled by inverse powers of $\h$ upon performing holographic renormalization. Interpreted as the number of degrees of freedom per site, we would conclude that this diverges as $\theta\rightarrow -\infty$, hinting at string-like dynamics. The connections discussed above may be a clue that ultrametricity plays a role in understanding these nonlocal theories, either in the effective holographic approach or in the full brane construction, the way it does in understanding nonlocal mean-field spin glasses. Such a growth of the number of degrees of freedom could analogously be true for the Q-space duals. It would then be very tempting to speculate that the sharpening of the ultrametric structure is holographically mirrored by the growth of the effective dimensionality of the dual, or alternatively the growth of the number of degrees of freedom per site, which are limits in which ultrametricity generically tends to sharpen  \cite{244332}.\\

\section*{Acknowledgements}
I would like to acknowledge useful conversations with  Frederik Denef, Xi Dong, Daniel Green, Sean Hartnoll, Shamit Kachru, Matt Kleban, Daniel Roberts, Subir Sachdev, Joshua Samani, Gonzalo Torroba, and especially Dionysios Anninos and Douglas Stanford. I am further indebted to Daniel Roberts and Douglas Stanford for sharing their computer code with me. I would also like to thank KU Leuven and the KITP for their warm hospitality while this work was in progress. I am supported by the Stanford Institute for Theoretical Physics under NSF Grant PHY-0756174.

\appendix
\begin{section}
{Form of higher curvature corrections}\label{proof2}
\end{section}
In this appendix we provide a short proof that adding higher curvature terms to an action which produces a hyperscaling-violating geometry leads to simple additions to the metric equation of motion. Specifically, for terms in the action consisting solely of contractions of the Riemann tensor, we will argue that their contribution to the effective stress energy tensor is of the form 
\be
T^{R^{n+1}}_{\mu\nu}\, dx^\mu dx^\nu=r^{-2n\h/d}\left(\alpha_n\frac{-dt^2}{r^{2z}}+\beta_n\frac{dr^2}{r^2}+\gamma_n\frac{dx_i^2}{r^2}\right).\label{provethis}
\ee
To prove this, it is important to notice that any entry of the Riemann tensor evaluated on the hyperscaling-violating metric ansatz (\ref{hyp2}) can be written as 
\be
R_{abcd}=r^{2\theta/d} f(d,z,\theta) r^{-2-2i} ,\label{riemann}
\ee
where $i$ equals either the number $1$ or the dynamical critical exponent $z$. Our point here is to isolate and focus on the $\theta$-dependence of any exponents of $r$. Now consider an arbitrary series of Riemann tensors before they are contracted:
\be
R_{abcd}R_{efgh}R_{ijkl}\cdots.
\ee
For $x$ Riemann tensors, there are $4x$ free indices and an overall $\theta$-dependent power of $r^{2x\theta/d}$. To contract all indices and produce a  scalar fit to be included in an action, we need $2x$ inverse metrics, which come in with a $\theta$-dependent power $r^{-4x\theta/d}$. The total $\theta$-dependent power of the produced curvature invariant is therefore $r^{-2x\theta/d}$. To produce the equations of motion one varies with respect to the inverse metric, which contributes a further power of $r^{2\theta/d}$, yielding $r^{-2(x-1)\theta/d}$. Tracking the measure of integration appropriately and identifying $x-1=n$, we see that at order $n+1$ in curvature the contribution to the effective stress energy tensor has an overall power $r^{-2n\theta/d}$. This explains the overall factor in (\ref{provethis}). Furthermore, since the constructed curvature invariant is, well, invariant, this must be the total power of $r$, i.e. the curvature invariant takes the form 
\be
(R_{ijkl})^{x}=r^{-2x\theta/d}f(z,\theta,d).
\ee
Varying with respect to the metric is what puts in the Lifshitz powers $r^{-2}$ and $r^{-2z}$ and explains the rest of the structure of (\ref{provethis}). For general scale covariance, i.e. a prefactor not of the form $r^{2\theta/d}$ but of arbitrary functional form $f(r)$, the structure of the Riemann tensor is not as simple as (\ref{riemann}) and the proof does not hold.

In the limit $\theta\rightarrow -\infty$ discussed in Section \ref{infinity}, we have the geometry 
\be
ds^2=e^{2r/\ell}\left(-dt^2+dr^2+dx_i^2\right).
\ee
Contributions to an effective stress-energy tensor of higher curvature corrections are analogous to the case above:
\be
T_{\mu\nu}^{R^{n+1}}dx^\mu dx^\nu=e^{-2rn/\ell}\left(-\alpha_n dt^2+\beta_n dr^2+ \gamma_n dx_i^2\right).
\ee
The proof mirrors the previous one.

\bibliographystyle{JHEP}

\begin{thebibliography}{99}
\bibitem{Maldacena:1997re} 
  J.~M.~Maldacena,
  Adv.\ Theor.\ Math.\ Phys.\  {\bf 2}, 231 (1998)
  [hep-th/9711200].
\bibitem{Hartnoll:2011fn} 
  S.~A.~Hartnoll,
  arXiv:1106.4324 [hep-th].
\bibitem{Kachru:2008yh} 
  S.~Kachru, X.~Liu and M.~Mulligan,
  Phys.\ Rev.\ D {\bf 78}, 106005 (2008)
  [arXiv:0808.1725 [hep-th]].
\bibitem{Son:2008ye} 
  D.~T.~Son,
  Phys.\ Rev.\ D {\bf 78}, 046003 (2008)
  [arXiv:0804.3972 [hep-th]].
\bibitem{Balasubramanian:2008dm} 
  K.~Balasubramanian and J.~McGreevy,
  Phys.\ Rev.\ Lett.\  {\bf 101}, 061601 (2008)
  [arXiv:0804.4053 [hep-th]].
\bibitem{Adams:2008zk} 
  A.~Adams, A.~Maloney, A.~Sinha and S.~E.~Vazquez,
  JHEP {\bf 0903}, 097 (2009)
  [arXiv:0812.0166 [hep-th]].
\bibitem{Anninos:2011kh} 
  D.~Anninos and F.~Denef,
  arXiv:1111.6061 [hep-th].
\bibitem{Huijse:2011ef} 
  L.~Huijse, S.~Sachdev and B.~Swingle,
  Phys.\ Rev.\ B {\bf 85}, 035121 (2012)
  [arXiv:1112.0573 [cond-mat.str-el]].
\bibitem{Gouteraux:2011ce} 
  B.~Gouteraux and E.~Kiritsis,
  JHEP {\bf 1112}, 036 (2011)
  [arXiv:1107.2116 [hep-th]].
\bibitem{Shaghoulian:2011aa} 
  E.~Shaghoulian,
  JHEP {\bf 1205}, 065 (2012)
  [arXiv:1112.2702 [hep-th]].
\bibitem{Copsey:2012gw} 
  K.~Copsey and R.~Mann,
  JHEP {\bf 1304}, 079 (2013)
  [arXiv:1210.1231 [hep-th]].
\bibitem{Horowitz:2011gh} 
  G.~T.~Horowitz and B.~Way,
  Phys.\ Rev.\ D {\bf 85}, 046008 (2012)
  [arXiv:1111.1243 [hep-th]].
\bibitem{Bao:2012yt} 
  N.~Bao, X.~Dong, S.~Harrison and E.~Silverstein,
  Phys.\ Rev.\ D {\bf 86}, 106008 (2012)
  [arXiv:1207.0171 [hep-th]].
\bibitem{Harrison:2012vy} 
  S.~Harrison, S.~Kachru and H.~Wang,
  arXiv:1202.6635 [hep-th].
\bibitem{Bhattacharya:2012zu} 
  J.~Bhattacharya, S.~Cremonini and A.~Sinkovics,
  JHEP {\bf 1302}, 147 (2013)
  [arXiv:1208.1752 [hep-th]].
\bibitem{Knodel:2013fua} 
  G.~Knodel and J.~T.~Liu,
  arXiv:1305.3279 [hep-th].
\bibitem{Ogawa:2011bz} 
  N.~Ogawa, T.~Takayanagi and T.~Ugajin,
  JHEP {\bf 1201}, 125 (2012)
  [arXiv:1111.1023 [hep-th]].
\bibitem{Ryu:2006bv} 
  S.~Ryu and T.~Takayanagi,
  Phys.\ Rev.\ Lett.\  {\bf 96}, 181602 (2006)
  [hep-th/0603001].
\bibitem{Wolf:2006zzb} 
  M.~M.~Wolf,
  Phys.\ Rev.\ Lett.\  {\bf 96}, 010404 (2006)
  [quant-ph/0503219].
\bibitem{Gioev:2006zz} 
  D.~Gioev and I.~Klich,
  Phys.\ Rev.\ Lett.\  {\bf 96}, 100503 (2006).
\bibitem{Hartnoll:2012wm} 
  S.~A.~Hartnoll and E.~Shaghoulian,
  JHEP {\bf 1207}, 078 (2012)
  [arXiv:1203.4236 [hep-th]].
\bibitem{Metlitski:2010pd} 
  M.~A.~Metlitski and S.~Sachdev,
  Phys.\ Rev.\ B {\bf 82}, 075127 (2010)
  [arXiv:1001.1153 [cond-mat.str-el]].
\bibitem{Thier:2011gf}
  S.~C.~Thier and W.~Metzner,
  Phys.\ Rev.\ B {\bf 84}, 155133 (2011)
  [arXiv:1108.1929 [cond-mat.str-el]]
\bibitem{kachru}
  S. Kachru, personal correspondence
\bibitem{Skenderis:2006jq} 
  K.~Skenderis and P.~K.~Townsend,
  Phys.\ Rev.\ Lett.\  {\bf 96}, 191301 (2006)
  [hep-th/0602260].
\bibitem{Skenderis:2006fb} 
  K.~Skenderis and P.~K.~Townsend,
  J.\ Phys.\ A {\bf 40}, 6733 (2007)
  [hep-th/0610253].
\bibitem{Skenderis:2007sm} 
  K.~Skenderis, P.~K.~Townsend and A.~Van Proeyen,
  JHEP {\bf 0708}, 036 (2007)
  [arXiv:0704.3918 [hep-th]].
\bibitem{Kiritsis:2013gia} 
  E.~Kiritsis,
  arXiv:1307.5873 [hep-th].
\bibitem{Sotiriou:2008rp} 
  T.~P.~Sotiriou and V.~Faraoni,
  Rev.\ Mod.\ Phys.\  {\bf 82}, 451 (2010)
  [arXiv:0805.1726 [gr-qc]].
\bibitem{DeFelice:2010aj} 
  A.~De Felice and S.~Tsujikawa,
  Living Rev.\ Rel.\  {\bf 13}, 3 (2010)
  [arXiv:1002.4928 [gr-qc]].
\bibitem{Nojiri:2006ri} 
  S.~'i.~Nojiri and S.~D.~Odintsov,
  eConf C {\bf 0602061}, 06 (2006)
  [Int.\ J.\ Geom.\ Meth.\ Mod.\ Phys.\  {\bf 4}, 115 (2007)]
  [hep-th/0601213].
\bibitem{Fay:2007uy} 
  S.~Fay, S.~Nesseris and L.~Perivolaropoulos,
  Phys.\ Rev.\ D {\bf 76}, 063504 (2007)
  [gr-qc/0703006 [GR-QC]].
\bibitem{Capozziello:2006dj} 
  S.~Capozziello, S.~Nojiri, S.~D.~Odintsov and A.~Troisi,
  Phys.\ Lett.\ B {\bf 639}, 135 (2006)
  [astro-ph/0604431].
\bibitem{Jaime:2012gc} 
  L.~G.~Jaime, L.~Patino and M.~Salgado,
  arXiv:1206.1642 [gr-qc].
\bibitem{Faraoni:2008mf} 
  V.~Faraoni,
  arXiv:0810.2602 [gr-qc].
\bibitem{GilMarin:2011xq} 
  H.~Gil-Marin, F.~Schmidt, W.~Hu, R.~Jimenez and L.~Verde,
  JCAP {\bf 1111}, 019 (2011)
  [arXiv:1109.2115 [astro-ph.CO]].
\bibitem{Nojiri:2003ft} 
  S.~'i.~Nojiri and S.~D.~Odintsov,
  Phys.\ Rev.\ D {\bf 68}, 123512 (2003)
  [hep-th/0307288].
\bibitem{Ohta:2004wk} 
  N.~Ohta,
  Int.\ J.\ Mod.\ Phys.\ A {\bf 20}, 1 (2005)
  [hep-th/0411230].
\bibitem{Maeda:2004hu} 
  K.~-i.~Maeda and N.~Ohta,
  Phys.\ Rev.\ D {\bf 71}, 063520 (2005)
  [hep-th/0411093].
\bibitem{Brevik:2006wa} 
  I.~H.~Brevik,
  Gen.\ Rel.\ Grav.\  {\bf 38}, 1317 (2006)
  [gr-qc/0603025].
\bibitem{Nojiri:2008nk} 
  S.~'i.~Nojiri and S.~D.~Odintsov,
  arXiv:0801.4843 [astro-ph].
\bibitem{Nojiri:2012zu} 
  S.~'i.~Nojiri and S.~D.~Odintsov,
  Phys.\ Lett.\ B {\bf 716}, 377 (2012)
  [arXiv:1207.5106 [hep-th]].
\bibitem{Hellerman:2001yi} 
  S.~Hellerman, N.~Kaloper and L.~Susskind,
  JHEP {\bf 0106}, 003 (2001)
  [hep-th/0104180].
\bibitem{Bousso:2004tv} 
  R.~Bousso,
  Phys.\ Rev.\ D {\bf 71}, 064024 (2005)
  [hep-th/0412197].
\bibitem{Hartle:1983ai} 
  J.~B.~Hartle and S.~W.~Hawking,
  Phys.\ Rev.\ D {\bf 28}, 2960 (1983).
\bibitem{Bunch:1978yq} 
  T.~S.~Bunch and P.~C.~W.~Davies,
  Proc.\ Roy.\ Soc.\ Lond.\ A {\bf 360}, 117 (1978).
\bibitem{Dong:2012se} 
  X.~Dong, S.~Harrison, S.~Kachru, G.~Torroba and H.~Wang,
  JHEP {\bf 1206}, 041 (2012)
  [arXiv:1201.1905 [hep-th]].
\bibitem{Ford:1977in} 
  L.~H.~Ford and L.~Parker,
  Phys.\ Rev.\ D {\bf 16}, 245 (1977).
\bibitem{Freedman:1991tk} 
  D.~Z.~Freedman, K.~Johnson and J.~I.~Latorre,
  Nucl.\ Phys.\ B {\bf 371}, 353 (1992).
\bibitem{Freedman:1992gr} 
  D.~Z.~Freedman, K.~Johnson, R.~Munoz-Tapia and X.~Vilasis-Cardona,
  Nucl.\ Phys.\ B {\bf 395}, 454 (1993)
  [hep-th/9206028].
\bibitem{Harlow:2011ke} 
  D.~Harlow and D.~Stanford,
  arXiv:1104.2621 [hep-th].
\bibitem{Anninos:2011ui} 
  D.~Anninos, T.~Hartman and A.~Strominger,
  arXiv:1108.5735 [hep-th].
\bibitem{Anninos:2012ft} 
  D.~Anninos, F.~Denef and D.~Harlow,
  arXiv:1207.5517 [hep-th].
\bibitem{Anninos:2013rza} 
  D.~Anninos, F.~Denef, G.~Konstantinidis and E.~Shaghoulian,
  arXiv:1305.6321 [hep-th].
\bibitem{Ng:2012xp} 
  G.~S.~Ng and A.~Strominger,
  Class.\ Quant.\ Grav.\  {\bf 30}, 104002 (2013)
  [arXiv:1204.1057 [hep-th]].
\bibitem{Das:2012dt} 
  D.~Das, S.~R.~Das, A.~Jevicki and Q.~Ye,
  JHEP {\bf 1301}, 107 (2013)
  [arXiv:1205.5776 [hep-th]].
\bibitem{Karch:2013oqa} 
  A.~Karch and C.~F.~Uhlemann,
  arXiv:1306.0582 [hep-th].
\bibitem{Banerjee:2013mca} 
  S.~Banerjee, A.~Belin, S.~Hellerman, A.~Lepage-Jutier, A.~Maloney, D.~Radicevic and S.~Shenker,
  arXiv:1306.6629 [hep-th].
\bibitem{Starobinsky:1982ee} 
  A.~A.~Starobinsky,
  Phys.\ Lett.\ B {\bf 117}, 175 (1982).
\bibitem{Strominger:2001pn} 
  A.~Strominger,
  JHEP {\bf 0110}, 034 (2001)
  [hep-th/0106113].
\bibitem{Strominger:2001gp} 
  A.~Strominger,
  JHEP {\bf 0111}, 049 (2001)
  [hep-th/0110087].
\bibitem{Witten:2001kn} 
  E.~Witten,
  hep-th/0106109.
\bibitem{Maldacena:2002vr} 
  J.~M.~Maldacena,
  JHEP {\bf 0305}, 013 (2003)
  [astro-ph/0210603].
\bibitem{Bousso:2002ju} 
  R.~Bousso,
  Rev.\ Mod.\ Phys.\  {\bf 74}, 825 (2002)
  [hep-th/0203101].
\bibitem{Bousso:1999dw} 
  R.~Bousso,
  Class.\ Quant.\ Grav.\  {\bf 17}, 997 (2000)
  [hep-th/9911002].
\bibitem{Bousso:1999xy} 
  R.~Bousso,
  JHEP {\bf 9907}, 004 (1999)
  [hep-th/9905177].
\bibitem{Brown:1992br} 
  J.~D.~Brown and J.~W.~York, Jr.,
  Phys.\ Rev.\ D {\bf 47}, 1407 (1993)
  [gr-qc/9209012].
\bibitem{Ammon:2012je} 
  M.~Ammon, M.~Kaminski and A.~Karch,
  JHEP {\bf 1211}, 028 (2012)
  [arXiv:1207.1726 [hep-th]].
\bibitem{Anninos:2012qw} 
  D.~Anninos,
  Int.\ J.\ Mod.\ Phys.\ A {\bf 27}, 1230013 (2012)
  [arXiv:1205.3855 [hep-th]].
\bibitem{Ade:2013zuv} 
  P.~A.~R.~Ade {\it et al.}  [Planck Collaboration],
  arXiv:1303.5076 [astro-ph.CO].
\bibitem{Jevicki:1998yr} 
  A.~Jevicki and T.~Yoneya,
  Nucl.\ Phys.\ B {\bf 535}, 335 (1998)
  [hep-th/9805069].
\bibitem{Jevicki:1998ub} 
  A.~Jevicki, Y.~Kazama and T.~Yoneya,
  Phys.\ Rev.\ D {\bf 59}, 066001 (1999)
  [hep-th/9810146].
\bibitem{Boonstra:1998mp} 
  H.~J.~Boonstra, K.~Skenderis and P.~K.~Townsend,
  JHEP {\bf 9901}, 003 (1999)
  [hep-th/9807137].
\bibitem{Kanitscheider:2008kd} 
  I.~Kanitscheider, K.~Skenderis and M.~Taylor,
  JHEP {\bf 0809}, 094 (2008)
  [arXiv:0807.3324 [hep-th]].
\bibitem{Harlow:2010my} 
  D.~Harlow and L.~Susskind,
  arXiv:1012.5302 [hep-th].
\bibitem{Anninos:2011jp} 
  D.~Anninos, G.~S.~Ng and A.~Strominger,
  JHEP {\bf 1202}, 032 (2012)
  [arXiv:1106.1175 [hep-th]].
\bibitem{spinglassbook}
  M.~Mezard, G.~Parisi, and M.~A.~Virasoro, \emph{Spin Glass Theory and Beyond.} vol. 9,
  \emph{Lecture Notes in Physics}, (World Scientic, 1987). ISBN 9971501155. \href{http:
  //books.google.com/books?id=ZIF9QgAACAAJ}{Google books}.
\bibitem{LNF-79-31-P} 
  G.~Parisi,
  ``An Infinite Number Of Order Parameters For Spin Glasses,''
  Phys.\ Rev.\ Lett.\ \ {\bf 43}, 1754  (1979).
\bibitem{parisi2} G.~Parisi, ``A sequence of approximated solutions to the SK model for spin glasses,'' J. Phys. A: Math. Gen. 13 L115 (1980).
\bibitem{198330} 
  G.~Parisi,
  ``Order parameter for spin-glasses,''
  Phys.\ Rev.\ Lett.\ \ {\bf 50}, 1946  (1983).
\bibitem{107911} 
  D.~Sherrington and S.~Kirkpatrick,
  ``Solvable Model of a Spin-Glass,''
  Phys.\ Rev.\ Lett.\ \ {\bf 35}, 1792  (1975).
\bibitem{244332} 
  R.~Rammal, G.~Toulouse and M.~A.~Virasoro,
  ``Ultrametricity for physicists,''
  Rev.\ Mod.\ Phys.\ \ {\bf 58}, 765  (1986).
\bibitem{Denef:2011ee}
  F.~Denef,
  ``TASI lectures on complex structures,''
  [arXiv:1104.0254 [hep-th]].
\bibitem{diothesis}
  D.~Anninos,
  ``Classical and Quantum Symmetries of De Sitter Space,"
  Ph.D Thesis, Harvard University (May 2011)
\bibitem{Benna:2011as} 
  M.~K.~Benna,
  Nucl.\ Phys.\ B {\bf 867}, 82 (2013)
  [arXiv:1111.4195 [hep-th]].
\bibitem{Roberts:2012jw} 
  D.~A.~Roberts and D.~Stanford,
  arXiv:1210.5238 [hep-th].
\bibitem{pedestrians}
T.~Castellani and A.~Cavagna,
J.\ Stat.\ Mech (2005) P05012
[arXiv: 0505032 [cond-mat.dis-nn]]
\bibitem{Gubser:2009qt} 
  S.~S.~Gubser and F.~D.~Rocha,
  Phys.\ Rev.\ D {\bf 81}, 046001 (2010)
  [arXiv:0911.2898 [hep-th]].
\bibitem{Anantua:2012nj} 
  R.~J.~Anantua, S.~A.~Hartnoll, V.~L.~Martin and D.~M.~Ramirez,
  JHEP {\bf 1303}, 104 (2013)
  [arXiv:1210.1590 [hep-th]].
\bibitem{Kutasov:2001uf} 
  D.~Kutasov,
  ``Introduction to little string theory.''
\bibitem{Ryu:2006ef} 
  S.~Ryu and T.~Takayanagi,
  JHEP {\bf 0608}, 045 (2006)
  [hep-th/0605073].
\bibitem{Kulaxizi:2012gy} 
  M.~Kulaxizi, A.~Parnachev and K.~Schalm,
  JHEP {\bf 1210}, 098 (2012)
  [arXiv:1208.2937 [hep-th]].
\bibitem{Barbon:2008ut} 
  J.~L.~F.~Barbon and C.~A.~Fuertes,
  JHEP {\bf 0804}, 096 (2008)
  [arXiv:0803.1928 [hep-th]].
\bibitem{Li:2010dr} 
  W.~Li and T.~Takayanagi,
  Phys.\ Rev.\ Lett.\  {\bf 106}, 141301 (2011)
  [arXiv:1010.3700 [hep-th]].
\bibitem{Fisher:1986zzb} 
  D.~S.~Fisher and D.~A.~Huse,
  Phys.\ Rev.\ Lett.\  {\bf 56}, 1601 (1986).
\bibitem{Belinsky:1970ew} 
  V.~A.~Belinsky, I.~M.~Khalatnikov and E.~M.~Lifshitz,
  Adv.\ Phys.\  {\bf 19}, 525 (1970).
\bibitem{Belinsky:1973:me}
  V.~A.~Belinsky and I.~M.~Khalatnikov 
  Sov.\ Phys.\ JETP {\bf 36}, 591-597 (1973).
\bibitem{Henneaux:2007ej} 
  M.~Henneaux, D.~Persson and P.~Spindel,
  Living Rev.\ Rel.\  {\bf 11}, 1 (2008)
  [arXiv:0710.1818 [hep-th]].
\bibitem{Ashtekar:2009vc} 
  A.~Ashtekar and E.~Wilson-Ewing,
  Phys.\ Rev.\ D {\bf 79}, 083535 (2009)
  [arXiv:0903.3397 [gr-qc]].
\bibitem{Korovin:2013bua} 
  Y.~Korovin, K.~Skenderis and M.~Taylor,
  arXiv:1304.7776 [hep-th].
\end{thebibliography}

\end{document}